\documentclass[fleqn,usenatbib]{mnras}

\usepackage{newtxtext,newtxmath}

\usepackage[T1]{fontenc}
\usepackage{ae,aecompl}

\usepackage{graphicx}
\usepackage[dvipsnames]{xcolor}
\usepackage{amsmath}
\usepackage{savesym}
\savesymbol{tablenum}
\usepackage{makecell}
\usepackage{siunitx}
\usepackage{xspace}
\usepackage{listings}
\usepackage{soul}
\usepackage{caption}
\restoresymbol{SIX}{tablenum}
\usepackage{bm}		

\usepackage{comment}

\usepackage{fontawesome}

\usepackage{xcolor,xspace}
\usepackage{booktabs}  
\usepackage{morefloats}  

\usepackage[switch,mathlines]{lineno}  
\usepackage{enumerate} 
\usepackage{tabularx}

\definecolor{linkcolor}{rgb}{0, 0, 1.}			
\definecolor{funcolor}{rgb}{0.65, 0.16, 0.16}	
\definecolor{parcolor}{HTML}{1E8449}        
\definecolor{bg_code}{rgb}{0.95,0.95,0.95}		

\newcommand{\tbref}[1]{Table~\ref{#1}}
\newcommand{\figref}[1]{Figure~\ref{#1}}

\newcommand{\alphamm}{\alpha_\mathrm{0.9-3.1\,mm}}

\renewcommand{\vec}[1]{\bm{#1}}

\newcommand{\amax}{a_\mathrm{max}}

\newcommand{\Lmm}{L_\mathrm{mm}}
\newcommand{\rhoeff}{\rho_\mathrm{eff}}
\newcommand{\Rse}{R_\mathrm{68}}
\newcommand{\rhose}{\rho_\mathrm{68}}

\newcommand{\ff}{\mathcal{F}}
\hypersetup{
draft=true,
colorlinks=true,
linkcolor=linkcolor,
citecolor=linkcolor,
filecolor=linkcolor,
urlcolor=linkcolor}

\hypersetup{
pdfauthor={M. Tazzari},
pdftitle={Multi-wavelength continuum sizes of protoplanetary discs},
pdfkeywords={pdf keywords},
bookmarksnumbered=true,
draft=false}

\title[Multi-wavelength continuum disc sizes]{
Multi-wavelength continuum sizes of protoplanetary discs:
scaling relations and implications for grain growth and radial drift
}

\author[M. Tazzari et al.]{M. Tazzari$^{1}$\thanks{Contact e-mail: \href{mailto:mtazzari@ast.cam.ac.uk}{mtazzari@ast.cam.ac.uk}},
C. J. Clarke$^{1}$, 
L. Testi$^{2,3}$,
J. P. Williams$^{4}$,
S. Facchini$^{2}$,
C. F. Manara$^{2}$,\newauthor
A. Natta$^{5}$
and
G. Rosotti$^{6,7}$
\\
$^{1}$Institute of Astronomy, University of Cambridge, Madingley Road, CB3 0HA,  Cambridge, UK\\
$^{2}$European Southern Observatory, Karl-Schwarzschild-Str. 2, D-85748 Garching, Germany\\
$^{3}$INAF/Osservatorio Astrofisico di Arcetri, Largo E. Fermi 5, I-50125 Firenze, Italy\\
$^{4}$Institute for Astronomy,University of Hawai‘i at Manoa,2680 Woodlawn Dr., Honolulu, HI, USA\\
$^{5}$School of Cosmic Physics, Dublin Institute for Advanced Studies, 31 Fitzwilliam Place, Dublin 2, Ireland\\
$^{6}$Leiden Observatory, Leiden University, P.O. Box 9531, NL-2300 RA Leiden, the Netherlands\\
$^{7}$School of Physics and Astronomy, University of Leicester, Leicester LE1 7RH, UK
}

\date{\today}

\pubyear{2020}

\begin{document}
\label{firstpage}
\pagerange{\pageref{firstpage}--\pageref{lastpage}}
\maketitle

\begin{abstract}
We analyse spatially resolved ALMA observations at 0.9, 1.3, and 3.1\,mm for the 26 brightest protoplanetary discs in the Lupus star-forming region. We characterise the discs multi-wavelength brightness profiles by fitting the interferometric visibilities in a homogeneous way, obtaining effective disc sizes at the three wavelengths, spectral index profiles and optical depth estimates.
We report three fundamental discoveries: first, the millimeter continuum size - luminosity relation already observed at 0.9\,mm is also present at 1.3\,mm with an identical slope, and at 3.1\,mm with a steeper slope, confirming that emission at longer wavelengths becomes increasingly optically thin. 
Second, when observed at 3.1\,mm the discs appear to be only 9\% smaller than when observed at 0.9\,mm, in tension with models of dust evolution which predict a starker difference.
Third, by forward modelling the sample of measurements with a simple parametric disc model, we find that the presence of large grains ($a_\mathrm{max}>1\,$mm) throughout the discs is the most favoured explanation for all discs as it reproduces simultaneously their spectral indices, optical depth, luminosity, and radial extent in the 0.9-1.3\,mm wavelength range. We also find that the observations can be alternatively interpreted with the discs being dominated by optically thick, unresolved, substructures made of mm-sized grains with a high scattering albedo. 
\end{abstract}




\section{Introduction}
\label{sect:introduction}
Multi-wavelength observations of protoplanetary discs at millimeter wavelengths have long been used to place constraints on the typical sizes of dust grains in discs \citep[and references therein]{Testi:2014kx}. 
This is important on several grounds. Firstly, if the grain size distribution is observationally calibrated then it improves estimates of the disc opacity;
provided the emission is optically thin, it improves estimates of the mass of solid materials in discs, a quantity of obvious interest when assessing the potential of discs to form rocky planets. 
Secondly, grain size determines the strength of coupling between the dust and gas phases; for typical gas densities and radii in the disc, grains of size around a mm are subject to strong radial drift towards the star \citep{Weidenschilling:1977lr}. Thus measuring the grain size allows one to assess how tightly the gas and dust dynamics are coupled and thus to determine whether the dust distribution is likely to be a good indicator of the underlying gas profile.

Nevertheless, the information that can be gathered from unresolved multi-wavelength observations is limited for a number of reasons. Not only is there the obvious problem of trying to derive mean characteristics from emission arising from a range of spatial locations, but, without
knowing the actual spatial extent of the emission, it is impossible to assess its optical depth. Ideally one would want to obtain well resolved radial profiles of disc emission at multiple mm wavelengths and hence derive local radial profiles of spectral index. This has however been performed only on few bright discs 
(AS~209 by \citealt{Perez:2012fk};
DoAr~25 and CY~Tau by \citealt{Perez:2015fk};
FT~Tau and DR~Tau by \citealt{Tazzari:2016qy}; UZ~Tau~E by \citealt{Tripathi:2018aa};
TW~Hya by \citealt{Huang:2018aa} and \citealt{Macias:2021aa};
HL~Tau by \citealt{Carrasco-Gonzalez:2019aa};
GM~Aur by \citealt{Huang:2020aa};
DS~Tau, GO~Tau, and DL~Tau by \citealt{Long:2020ab}),
which makes it hard to derive statistical properties of disc populations.

An intermediate situation with regard to sample size and resolution is attained in the case of our ALMA surveys of the Lupus protoplanetary discs, for which we have gathered observations at intermediate resolution (0.25-0.35$\arcsec$) across the 0.9-3.1\,mm wavelength range for 36 objects.
In \citet{Tazzari:2020aa} we presented the 3\,mm survey (with particular focus on the spatially-integrated fluxes and spectral indices), which complements the 0.9\,mm \citep{Ansdell:2016qf} and 1.3\,mm \citep{Ansdell:2018aa} surveys. 
Lupus is so far the only star forming region that has been targeted with ALMA survey-like observations at three wavelengths: the homogeneity of the data and the statistical relevance of the targeted sample make it the ideal benchmark to test our understanding of dust evolution and disc structure.

With an average distance from us of 160\,pc \citep{Manara:2018aa}, Lupus discs are partially resolved with typically 4-6 resolution elements across the disc diameter.
While it is possible to derive spectral index profiles from such data, this typically extends over a limited radial range and lacks information on small scale
structure.
Notwithstanding, a quantity that can be robustly derived from even moderate resolution data is the disc radius (practically, the radius enclosing $68\%$ of the flux, $\Rse$) at different wavelengths. 
The ratio of such radii at different wavelengths turns out to be a rather constraining quantity. For example, in the case of a disc where the maximum grain size decreases rapidly with radius, the less efficient radiation of small grains at long wavelengths would lead to the disc size being smaller as the wavelength of observation increases. On the other hand, if the entire disc is optically thick, or if the maximum grain size is large enough at all radii so that the disc emits efficiently at all the wavebands considered, the disc size
would be expected to depend more weakly on radius.

Another quantity that can be assessed with data that is moderately resolved, as in the case of Lupus discs, is how close the disc is to being optically thick.
Here we define the optically thick fraction as the ratio of the observed flux
within $\Rse$ to the flux of a disc of size $\Rse$ where the emission was completely optically thick (having made an estimate of the likely temperature profile of the disc according to the stellar luminosity). We find that, in conjunction, the spatially averaged spectral indices \citep{Tazzari:2020aa}, 
the ratio of disc sizes in the three ALMA wavebands, and the optically thick fraction provides information that is quite constraining of the required disc properties, even in cases where there is very little direct information about the radial profiles of disc emission.

In this paper we analyse spatially resolved ALMA observations at 0.9, 1.3, and 3.1\,mm for the 26 brightest protoplanetary discs in the Lupus star-forming region. We characterise the discs multi-wavelength brightness profiles by fitting the interferometric visibilities in a homogeneous way, obtaining effective disc sizes at the three wavelengths, spectral index profiles and optical depth estimates.

The paper is structured as follows.
In Section~\ref{sect:sample} we present the sample of young stellar objects that we consider.
Section~\ref{sect:analysis} describes how the visibility modelling is performed and Section~\ref{sect:results.fit.example} presents a worked example for one disc.
Section~\ref{sect:results.demographic.properties} presents the results of the modelling in terms of demographic properties (disc radii, optical depth, continuum size-luminosity scaling relation). 
In Section~\ref{sect:discussion} we employ simple toy models of the disc emission to interpret the sample of measurements in terms of disc structure and dust properties, discussing the implications for radial drift and grain growth.
Finally, in Section~\ref{sect:conclusions} we draw our conclusions. 
Appendix~\ref{app:size.luminosity.inc.correction} presents the linear regression of the millimeter size-integrated flux relation.
Appendices~\ref{app:detailed.toymodel.results:smooth}  and \ref{app:detailed.toymodel.results:structured} show detailed properties of some toy models representative of different regimes.
Appendix~\ref{app:role.of.scattering} discusses the effects of scattering on the inferred dust properties.
Appendix~\ref{app:detailed.fit.results} reports the detailed multi-wavelength fit results for all the discs.

\section{Sample and Observations}
\label{sect:sample}
Protoplanetary discs in the Lupus star forming region have been recently targeted with extensive ALMA surveys at moderate sensitivity and resolution (0.25$\arcsec$-0.35$\arcsec$) that provided the first systematic view on the structure of these discs. Here we assemble all the data into a multi-wavelength dataset collecting 0.9\,mm observations (ALMA Band~7) from \citet{Ansdell:2016qf}, 1.3\,mm observations (ALMA Band~6) from \citet{Ansdell:2018aa}, and 3.1\,mm observations (ALMA Band~3) from Tazzari et al. (2020) for all the discs in common to these surveys. 
The targets originally selected by these ALMA surveys are sources typically classified as Class~II discs \citep{Merin:2008xy} or with a flat infrared excess measured between the 2MASS Ks (2.2$\mu$m) and \textit{Spitzer} MIPS-1 (24$\mu$m) bands \citep{Evans:2009ys}. Please refer to the survey papers for full details on the sample selection.
Note that IM~Lup (Sz~82) was not targeted at 0.9\,mm by \citet{Ansdell:2016qf}: we thus use the 0.9\,mm observations at similar sensitivity and resolution by \citet{Cleeves:2016lr}. 

The combined sample of these three surveys results in 35 sources targeted at 0.9, 1.3, and 3.1\,mm. This constitutes the initial sample for the present study. Since Sz~91 has not been detected at 3.1\,mm \citep{Tazzari:2020aa} we remove it from the sample. Moreover, we want to consider only discs that orbit around single stars or wide orbit binary companions, in order to avoid biases in the disc properties induced by the presence of the companion (e.g., tidal truncation effects; see \citealt{Manara:2019aa,Akeson:2019aa}).  Since the interferometric visibilities contain contributions from all the sources in the field of view, analysing discs in multiple systems requires fitting all their components simultaneously \citep[e.g.,][]{Manara:2019aa}, which is outside the scope of this paper. We thus exclude two close binaries (Sz~74, and V856~Sco/Lupus~III~53) and a triple system (Sz~68/HT~Lup) with separations less than $2\arcsec$.
The resulting sample 
is thus made of 31 sources: their coordinates and properties are summarised in \tbref{tb:YSOs}. 
We note that we will remove 5 of these sources from the sample for different reasons (see Section~\ref{sect:modelling.special.cases}), leaving 26 sources for the full multi-wavelength analysis.
\begin{table*}
 \caption{Properties of the Lupus young stellar objects  considered for the multi-wavelength analysis.}
 \centering
 \begin{tabular}{llllccccl}
   \midrule
   \toprule
Name
 & Other Name
 & R.A. & Dec.
 & 2MASS identifier & Gaia DR2 identifier
 & $d$ & $L_{\star}$ & Notes \\ 

 & & (J2015.5) & (J2015.5)
 &  & 
 & (pc) & ($L_\odot$) &  \\ 
   \midrule
Sz 65               
 &                     
 & 15:39:27.76    & -34:46:17.55  
 & J15392776-3446171    &  6013399894569703040
 & $155$ & $0.89$
 &  \\ 
Sz 66               
 &                     
 & 15:39:28.27    & -34:46:18.42  
 & J15392828-3446180    &  6013399830146943104
 & $157$ & $0.20$
 & $^{(3)}$ \\ 
J15450634-3417378   
 &                     
 & 15:45:06.34    & -34:17:37.83  
 & J15450634-3417378    & ...                 
 & $160$ & ... & $^{(1)}$ $^{(3)}$ \\ 
J15450887-3417333   
 &                     
 & 15:45:08.86    & -34:17:33.80  
 & J15450887-3417333    &  6014696875913435520
 & $155$ & $0.06$
 &  \\ 
Sz 69               
 &                     
 & 15:45:17.39    & -34:18:28.64  
 & J15451741-3418283    &  6014696635395266304
 & $154$ & $0.09$
 &  \\ 
Sz 71               
 & GW Lup              
 & 15:46:44.71    & -34:30:36.04  
 & J15464473-3430354    &  6014722194741392512
 & $155$ & $0.33$
 & $^{(2)}$ \\ 
Sz 72               
 &                     
 & 15:47:50.61    & -35:28:35.76  
 & J15475062-3528353    &  6011573266459331072
 & $155$ & $0.27$
 &  \\ 
Sz 73               
 &                     
 & 15:47:56.93    & -35:14:35.14  
 & J15475693-3514346    &  6011593641784262400
 & $156$ & $0.42$
 & $^{(3)}$ \\ 
Sz 82               
 & IM Lup              
 & 15:56:09.19    & -37:56:06.49  
 & J15560921-3756057    &  6010135758090335232
 & $158$ & $2.60$
 & $^{(2)}$ \\ 
Sz 83               
 & RU Lup              
 & 15:56:42.30    & -37:49:15.83  
 & J15564230-3749154    &  6010114558131195392
 & $159$ & $1.49$
 & $^{(2)}$ \\ 
Sz 84               
 &                     
 & 15:58:02.50    & -37:36:03.09  
 & J15580252-3736026    &  6010216537834709760
 & $152$ & $0.13$
 &  \\ 
Sz 129              
 &                     
 & 15:59:16.46    & -41:57:10.66  
 & J15591647-4157102    &  5995168724780802944
 & $161$ & $0.43$
 & $^{(2)}$ \\ 
J15592838-4021513   
 & RY Lup              
 & 15:59:28.37    & -40:21:51.59  
 & J15592838-4021513    &  5996151172781298304
 & $158$ & $1.87$
 &  \\ 
J16000236-4222145   
 &                     
 & 16:00:02.34    & -42:22:14.96  
 & J16000236-4222145    &  5995139484643284864
 & $163$ & $0.18$
 &  \\ 
J16004452-4155310   
 & MY Lup              
 & 16:00:44.50    & -41:55:31.29  
 & J16004452-4155310    &  5995177933191206016
 & $156$ & $0.85$
 & $^{(2)}$ \\ 
J16011549-4152351   
 &                     
 & 16:01:15.49    & -41:52:35.19  
 & J16011549-4152351    & ...                 
 & $160$ & ... & $^{(1)}$ \\ 
Sz 133              
 &                     
 & 16:03:29.37    & -41:40:02.17  
 & J16032939-4140018    &  5995094095435598848
 & $155$ & $0.07$
 &  \\ 
J16070854-3914075   
 &                     
 & 16:07:08.54    & -39:14:07.89  
 & J16070854-3914075    &  5997076721058575360
 & $177$ & $0.14$
 &  \\ 
Sz 90               
 &                     
 & 16:07:10.06    & -39:11:03.65  
 & J16071007-3911033    &  5997077167735183872
 & $160$ & $0.42$
 &  \\ 
Sz 98               
 & HK Lup              
 & 16:08:22.48    & -39:04:46.81  
 & J16082249-3904464    &  5997082867132347136
 & $156$ & $1.53$
 &  \\ 
Sz 100              
 &                     
 & 16:08:25.75    & -39:06:01.59  
 & J16082576-3906011    &  5997082046818385408
 & $137$ & $0.08$
 &  \\ 
J16083070-3828268   
 &                     
 & 16:08:30.69    & -38:28:27.24  
 & J16083070-3828268    &  5997490206145065088
 & $155$ & $1.84$
 & $^{(1)}$ \\ 
Sz 108B             
 &                     
 & 16:08:42.87    & -39:06:15.03  
 & ...                  &  5997082218616859264
 & $168$ & $0.11$
 &  \\ 
Sz 110              
 &                     
 & 16:08:51.56    & -39:03:18.07  
 & J16085157-3903177    &  5997082390415552768
 & $159$ & $0.17$
 & $^{(3)}$ \\ 
J16085324-3914401   
 &                     
 & 16:08:53.23    & -39:14:40.53  
 & J16085324-3914401    &  5997033290348155136
 & $167$ & $0.21$
 &  \\ 
Sz 111              
 &                     
 & 16:08:54.67    & -39:37:43.50  
 & J16085468-3937431    &  5997006897751436544
 & $158$ & $0.21$
 &  \\ 
Sz 113              
 &                     
 & 16:08:57.79    & -39:02:23.21  
 & J16085780-3902227    &  5997457736191421184
 & $163$ & $0.06$
 & $^{(3)}$ \\ 
Sz 114              
 &                     
 & 16:09:01.84    & -39:05:12.79  
 & J16090185-3905124    &  5997410491550194816
 & $162$ & $0.21$
 & $^{(2)}$ \\ 
Sz 118              
 &                     
 & 16:09:48.64    & -39:11:17.21  
 & J16094864-3911169    &  5997405509388068352
 & $163$ & $0.72$
 &  \\ 
Sz 123              
 &                     
 & 16:10:51.57    & -38:53:14.13  
 & J16105158-3853137    &  5997416573223873536
 & $162$ & $0.13$
 &  \\ 
J16124373-3815031   
 &                     
 & 16:12:43.74    & -38:15:03.42  
 & J16124373-3815031    &  5997549820286701440
 & $159$ & $0.39$
 &  \\
   \bottomrule
 \end{tabular}
 \begin{flushleft}
 \textbf{Note.}
 Name is the designation used in this paper (with notable alternative names where available).
 Right Ascension (R.A.) and Declination (Dec.) are from Gaia DR2 \citep{Gaia-Collaboration:2018aa}. 
 Distances ($d$) are computed using Gaia DR2 data by \cite{Bailer-Jones:2018aa}. 
 Stellar bolometric luminosities ($L_\star$) are measured by \cite{Alcala:2017aa} from spectroscopic observations and corrected for Gaia DR2 distances by \cite{Alcala:2019aa}.
$^{(1)}$~Not found in Gaia DR2. For this source, we assume the average distance of the Lupus cloud complex (160 pc, \citealt{Manara:2018aa}).
$^{(2)}$~Observed in the DSHARP survey \citep{Andrews:2018aa}. 
$^{(3)}$~Sources that will be removed from the multi-wavelength analysis for different reasons (see Sect.~\ref{sect:modelling.special.cases}). 
A machine-readable version of this table is available online (see the Data Availability statement).
\end{flushleft}
\label{tb:YSOs}
\end{table*}

\section{Analysis}
\label{sect:analysis}

\subsection{Modelling the disc brightness profiles}
To characterise the surface brightness profile of all discs we analyse the ALMA observations at all three  observing wavelengths (0.89, 1.3, and 3.1\,mm) in a homogeneous way. We fit the complex visibilities with a parametric brightness profile under the assumption that the disc emission can be represented well by an axisymmetric profile. At the resolution of these observations this assumption is typically well satisfied a posteriori, as the fit residuals confirm in all but a few cases (cf. detailed fit results in Appendix~\ref{app:detailed.fit.results}).
Inferring the brightness profile by fitting the visibilities (as opposed to extracting it from the corresponding synthesized image) allows for a more robust understanding of the uncertainties and for a spatial resolution typically better than the nominal synthesized beam.

Past studies adopted different parametrisations to fit protoplanetary discs brightness profiles: a sharply truncated power law \citep[e.g.,][]{Isella:2009qy, Guilloteau:2011aa}, the self-similar solution to the viscous evolution equation by \citealt{Lynden-Bell:1974aa} \citep[e.g., ][]{Andrews:2009zr, Isella:2009qy,Guilloteau:2011aa,Tazzari:2017aa}, and more recently the profile introduced by \citet{Lauer:1995aa} \citep[e.g., ][]{Tripathi:2017aa, Andrews:2018ab, Hendler:2020aa}. These three parametrisations have an increasing 
degree of flexibility, with the Nuker profile being also the most general one (see \citealt{Tripathi:2017aa} for a discussion of its properties). 
The additional flexibility comes with an increased number of parameters (5, compared to 3 for the former two profiles) that allow for a decoupled description of the inner and outer disc region. The studies that employed the Nuker profile analysed disc observations with resolution and sensitivity comparable or identical to those of the observations we analyse here and they all show that at least one of the parameters ($\alpha$, which controls  the sharpness of the transition between inner and outer disc) is hardly constrained by them. 
In this study we adopt a modified version of the self-similar profile that has 4 parameters and, although lacks Nuker's capability to reproduce sharp broken power-law profiles, is still very flexible and retains the advantageous decoupling between inner and outer disc description:
\begin{equation}
\label{eq:brightness.parametrisation}
    I_\nu(\rho)\propto \left(\frac{\rho}{\rho_c}\right)^{\gamma_1} \exp\left[-\left(\frac{\rho}{\rho_c}\right)^{\gamma_2}\right]\,,
\end{equation}
where $\rho_c$ ($\arcsec$) is the scaling radius, $\gamma_1$ is the power-law slope in the inner disc, and $\gamma_2$ controls the slope in the exponentially tapered outer disc. Note that if $\gamma_2=2-\gamma_1$, then Eq.~\eqref{eq:brightness.parametrisation} corresponds to the canonical self-similar solution and in the case of $\gamma_1=0,\ \gamma_2=2$ it reduces to the Gaussian. This profile has already been used successfully by \citet{Long:2019aa,Manara:2019aa}.
The normalisation of $I(\rho)$ is determined so that the integrated flux $F_\nu$ (mJy) matches the observations. We define the cumulative flux
\begin{equation}
\label{eq:def.cumulative.flux}
    f(\rho) = 2 \pi \int_\mathrm{0}^{\rho} I(\rho') \rho' d\rho'\,,
\end{equation}
where the inclination $i$ and brightness $I(\rho)$ have been inferred with the MCMC visibility modelling. By definition the integrated disc flux is ${F_\nu=f(\infty)}$.

The functional form in Eq.~\eqref{eq:brightness.parametrisation} has 4 free parameters: $F_\nu$, $\rho_c$, $\gamma_1$, and $\gamma_2$. 
Additionally, we fit the disc inclination $i$  along the line of sight, the position angle $PA$ (defined East of North), and the angular offsets $(\Delta$R.A.,$\Delta$Dec.$)$ from the phase center. We thus have an 8-dimensional parameter space
\begin{equation}
    \vec{\theta}=(F_\nu,\rho_c,\gamma_1,\gamma_2, i, PA, \Delta\mathrm{R.A.}, \Delta\mathrm{Dec.})\,,
\end{equation}
described by 4 parameters for the brightness profile plus 4 for the system geometry.
We compare a model $\vec\theta$ to the observations assuming a Gaussian likelihood
\begin{equation}
\log{}p(V_\mathrm{obs}\,|\,\vec{\theta})=-\frac{1}{2}\,\chi^2=-\frac{1}{2}\sum_{j=1}^{N}|V_\mathrm{obs,j}-V_\mathrm{mod,j}(\vec{\theta})|^2w_j\,,
\end{equation}
where $V_\mathrm{obs}$ are the observed complex visibilities, $V_\mathrm{mod}(\vec{\theta})$ are the model visibilities, $N$ is the total number of visibility points and $w_j$ is the weight\footnote{Theoretical visibility weights $w_j$ are computed by the CASA software package assuming Gaussian uncertainties \citep{Wrobel:1999gf}.} of the $j-$th visibility. We denote $V_\mathrm{j}=V(u_j, v_j)$ where $(u_j,v_j)$ is the Fourier-plane coordinate of the $j-$th visibility point.

For each model $\vec{\theta}$ we compute the visibilities $V_\mathrm{mod}$ using the \texttt{GALARIO} Python package\footnote{Code available at \href{https://github.com/mtazzari/galario}{https://github.com/mtazzari/galario}.} \citep{Tazzari:2018aa}, which first computes the 2D image of the disc for a given brightness $I(R)$ and then samples its Fourier transform at the observed $(u,v)$ locations.

We explore the parameter space with a Bayesian approach using an affine-invariant Markov chain Monte Carlo (MCMC) ensemble sampler implemented in the Python package \textsc{emcee} v2.2.1 \citep{2013PASP..125..306F}. We thus obtain samples of the posterior probability of the model parameters given the observations:
\begin{equation}
\log p(\vec\theta\,|\,V_\mathrm{obs}){}={}\log{}p(V_\mathrm{obs}\,|\,\vec\theta)+\log{}p(\vec\theta){}+{}C\,,
\end{equation}
where $p(\vec\theta)$ is the prior probability on the parameters and $C$ is a normalisation constant that can be neglected for the purposes of this study.
Since the parameters are independent, the priors can be factored as $p(\vec\theta)=\prod_i p(\theta_i)$.
We adopt uniform priors $U(a,b)$ between $a$ and $b$ for all parameters, 
$p(\log F_\nu/\mathrm{mJy})=
U(-3, 5)$,
$p(R_c)=
U(0, 5\arcsec)$,
$p(\gamma_1)=
U(-7, 7)$,
$p(\gamma_2)=
U(-7, 7)$,
$p(P.A.)=
U(0, 180^{\circ})$,
$p(\Delta\mathrm{RA})=
U(-5\arcsec, 5\arcsec)$,
$p(\Delta\mathrm{Dec})=
U(-5\arcsec, 5\arcsec)$,
while for the inclination we adopt $p(i)=\sin(i)$ for  $0\leq{}i\leq\pi/2$.

For all fits we use 48 walkers (i.e., 6 chains for each free parameter) that we initially draw from the same uniform distributions that we use as priors. We then evolve the walkers for $10^5$ steps, with convergence typically being achieved in the first 15-20 thousand steps (burn-in phase). 
To obtain independent samples of the posterior, we remove the burn-in steps from the entire $10^5$-steps long chain and we further \textit{thin} the remaining samples by selecting only steps separated by one autocorrelation time (typically in the range 100-150 steps for all discs). 

To prepare the observational data for the fits, we use the CASA \texttt{split} command to average the visibilities in spectrum (reducing the data to 1 channel per spectral window), and in time (over 30 seconds). By applying the CASA \texttt{statwt} command we also make sure that the absolute scale of the visibility weights is correct.

\subsection{Measuring the disc \textit{size}}
\label{sect:measuring.disc.size}
The disc size is not quantified directly by the modelling presented above.  Historically, studies that followed a similar approach of fitting the interferometric visibilities adopted the scaling radius associated with their parametrisation as the disc \textit{size}. In the case of the functional form that we choose here (Eq.~\ref{eq:brightness.parametrisation}) this would translate into taking $R_c$ as the disc size. As pointed out by \citet{Tripathi:2017aa}, the choice of a \textit{punctual} quantity such as a scaling radius is problematic for several reasons. 
A more robust way of measuring the disc size is through the definition of an effective radius $\rhoeff$ enclosing a fraction $x$ of the total disc emission (such that $f(\rhoeff)=xF_\nu$), which accounts for the limited sensitivity and resolution of the data and is much less sensitive to the degeneracies between the parameters. We note that with this definition of disc size, the particular functional form chosen to fit the brightness profile becomes less important: any parametrisation that reproduces the data well will suggest the same effective size \citep{Andrews:2018ab}. 

To ease the comparison with other studies, and given its immediate connection to a Gaussian standard deviation in the case of poorly resolved observations, throughout this study we use $x=0.68$: we will refer to the effective angular radius with $\rhose$ and with $\Rse=\rhose \times d$ to its linear counterpart ($d$ being the Gaia DR2 distance, see \tbref{tb:YSOs}).
Similarly to \citet{Tripathi:2017aa}, we report that choosing $x$ in the range between 0.5 and 0.8 has a negligible impact on the general conclusions of this work.

\subsection{Modified modelling for special cases}
\label{sect:modelling.special.cases}
There are 7 discs (J15450887-3417333, Sz 69, Sz 72, Sz 73, Sz 90, Sz 108B, Sz 110, J16085324-3914401, J16124373-3815031) for which the MCMC does not converge at one or more wavelengths. They are the faintest objects  in the sample across the three wavelengths, with $F_\nu <= 3$\,mJy at 3.1mm. They appear unresolved in the CLEAN-synthesized images (Briggs weighting, robust 0.5) but the deprojected visibilities show that they are resolved, albeit with a larger uncertainty given the noisier data. The MCMC posteriors for the modified self-similar profile in Eq.~\eqref{eq:brightness.parametrisation} are highly degenerate and tend to prefer very steep profiles ($\gamma_1<-2,\ \gamma_2>5$), with extremely large scaling radii $\rho_c \gg 5\arcsec$). Although it is possible that the underlying brightness of these sources is genuinely steeper than other discs, we found that equally good, and sometimes even better, solutions were found by adopting a Gaussian profile centered on the origin:
\begin{equation}
\label{eq:brigthness.parametrisation.gaussian}
    I_\nu(\rho) \propto \exp\left(\frac{-\rho^2}{2\sigma^2}\right)\,,
\end{equation}
where the normalisation (such that the integrated flux matches the observations) and the width $\sigma$ ($\arcsec$) are the only free parameters in the MCMC. In these cases we adopt uniform priors $p(\log(F_\nu/\mathrm{mJy})=U(-3,5)$ and $p(\sigma)=U(0,5\arcsec)$.

There are 4 discs (J15450634-3417378, Sz 73, Sz 110, Sz 113) with very noisy observations for which neither the modified self-similar nor the Gaussian profiles converge at one or more wavelengths. We thus exclude them from the analysis.

In our sample there is one binary system that has both the components detected at the three wavelengths: the wide binary Sz 65+Sz 66 (separation 6.4$\arcsec$).
Sz 65+Sz 66 were targeted at all three wavelengths with two distinct scheduling blocks, each centered on one of the components. Given their large separation, Sz 65 and Sz 66 fall in the respective primary beam of both the scheduling blocks. For Sz 65 we fit the visibility data from the scheduling block centered on it, after the Sz 66 contribution to the visibilities was removed with CASA by subtracting the Fourier transform of its CLEAN components.
Since Sz 66 is considerably fainter than Sz 65, the removal of Sz~65 emission from its scheduling block proved difficult. For this reason we do not report fit results for Sz~66. Out of the sample of 31 discs initially selected for the multi-wavelength analysis (Sect.~\ref{sect:sample}), we are therefore left with 26 of them.

\section{Results}
\label{sect:results.fit.example}
We fit the ALMA observations of 26 discs at 0.89, 1.3, and 3.1\,mm using the modified self-similar parametrisation in Eq.~\eqref{eq:brightness.parametrisation} for 19 of them, and the Gaussian parametrisation in Eq.~\eqref{eq:brigthness.parametrisation.gaussian} for 7 of them. Table~\ref{tb:fit.results.ss2s} summarises the statistics of the posterior distributions inferred for the free parameters of the modified self-similar brightness profile. Table~\ref{tb:fit.results.gaussian} presents analogous results for the Gaussian fits. The tables provide also the posteriors of derived quantities such as the integrated flux $F_\nu$, the effective (angular) radius $\rhose$, the effective linear radius $\Rse$, and the optically thick fraction $\mathcal{F}$ as defined in Sect.~\ref{sect:opt.depth.constraints}. 

\begin{table*}
 \caption{Parameters of the inferred brightness profiles: modified self-similar brightness.}
 \centering
 \begin{tabular}{lccccccc|ccc}
   \midrule
   \toprule
Name & $\lambda$
 & $F_\mathrm{\nu}$ & $\rho_c$ & $\gamma_1$ & $\gamma_2$
 & $i$ & $P.A.$
 & $\rho_{68}$ & $R_{68}$ & $\mathcal{F}$ \\ 
 & (mm)
 & (mJy) & ($\arcsec$) &  & 
 & $(^\circ)$ & $(^\circ)$
 & $(\arcsec)$ & (au) \\ 
   \midrule
Sz 65                & $0.89$
 & $63.5^{\pm0.8}$ & $0.18^{\pm0.05}$ & $-0.34^{\pm0.41}$ & $2.44^{\pm1.08}$
 & $56^{\pm1}$ & $109^{\pm1}$
 & $0.162^{\pm0.005}$ & $25.1^{\pm0.8}$ & $0.66$ \\ 
                     & $1.33$
 & $28.9^{\pm0.2}$ & $0.22^{\pm0.02}$ & $-0.56^{\pm0.14}$ & $3.94^{\pm0.86}$
 & $61^{\pm1}$ & $113^{\pm1}$
 & $0.166^{\pm0.003}$ & $25.7^{\pm0.5}$ & $0.67$ \\ 
                     & $3.10$
 & $4.9^{\pm0.2}$ & $0.24^{\pm0.09}$ & $-1.14^{\pm0.38}$ & $2.42^{\pm1.24}$
 & $47^{\pm6}$ & $100^{\pm7}$
 & $0.145^{\pm0.012}$ & $22.4^{\pm1.8}$ & $0.46$ \\ 
Sz 71                & $0.89$
 & $183.0^{\pm1.2}$ & $0.53^{\pm0.02}$ & $-0.66^{\pm0.04}$ & $1.96^{\pm0.13}$
 & $42^{\pm1}$ & $38^{\pm1}$
 & $0.450^{\pm0.004}$ & $69.9^{\pm0.7}$ & $0.56$ \\ 
                     & $1.33$
 & $73.9^{\pm0.5}$ & $0.45^{\pm0.03}$ & $-0.63^{\pm0.04}$ & $1.92^{\pm0.15}$
 & $40^{\pm1}$ & $27^{\pm1}$
 & $0.392^{\pm0.004}$ & $60.8^{\pm0.6}$ & $0.46$ \\ 
                     & $3.10$
 & $9.6^{\pm0.2}$ & $0.21^{\pm0.10}$ & $-0.46^{\pm0.24}$ & $1.08^{\pm0.25}$
 & $39^{\pm3}$ & $29^{\pm4}$
 & $0.339^{\pm0.011}$ & $52.6^{\pm1.7}$ & $0.30$ \\ 
Sz 82                & $0.90$
 & $674.8^{\pm0.8}$ & $1.96^{\pm0.01}$ & $-1.11^{\pm0.01}$ & $5.00^{\pm0.01}$
 & $52^{\pm1}$ & $144^{\pm1}$
 & $1.184^{\pm0.002}$ & $186.8^{\pm0.2}$ & $0.35$ \\ 
                     & $1.28$
 & $221.5^{\pm2.2}$ & $1.67^{\pm0.03}$ & $-1.27^{\pm0.01}$ & $4.52^{\pm0.32}$
 & $33^{\pm1}$ & $0^{\pm1}$
 & $0.905^{\pm0.013}$ & $142.8^{\pm2.0}$ & $0.20$ \\ 
                     & $3.10$
 & $19.6^{\pm0.2}$ & $0.00^{\pm0.01}$ & $-1.27^{\pm0.03}$ & $0.21^{\pm0.01}$
 & $26^{\pm1}$ & $0^{\pm1}$
 & $0.821^{\pm0.036}$ & $129.4^{\pm5.7}$ & $0.08$ \\ 
Sz 83                & $0.89$
 & $428.1^{\pm2.3}$ & $0.39^{\pm0.01}$ & $-0.73^{\pm0.05}$ & $3.73^{\pm0.51}$
 & $22^{\pm1}$ & $89^{\pm4}$
 & $0.279^{\pm0.002}$ & $44.2^{\pm0.3}$ & $1.12$ \\ 
                     & $1.33$
 & $167.3^{\pm0.7}$ & $0.39^{\pm0.01}$ & $-0.78^{\pm0.02}$ & $4.66^{\pm0.28}$
 & $22^{\pm1}$ & $125^{\pm3}$
 & $0.272^{\pm0.002}$ & $43.2^{\pm0.3}$ & $0.87$ \\ 
                     & $3.11$
 & $22.3^{\pm0.1}$ & $0.36^{\pm0.01}$ & $-0.98^{\pm0.03}$ & $4.38^{\pm0.45}$
 & $17^{\pm2}$ & $111^{\pm7}$
 & $0.233^{\pm0.002}$ & $37.1^{\pm0.3}$ & $0.66$ \\ 
Sz 84                & $0.89$
 & $32.3^{\pm0.5}$ & $0.12^{\pm0.07}$ & $1.29^{\pm0.67}$ & $1.40^{\pm0.55}$
 & $70^{\pm2}$ & $0^{\pm1}$
 & $0.236^{\pm0.006}$ & $35.9^{\pm0.9}$ & $0.67$ \\ 
                     & $1.33$
 & $12.6^{\pm0.2}$ & $0.17^{\pm0.07}$ & $0.55^{\pm0.51}$ & $1.63^{\pm0.49}$
 & $66^{\pm1}$ & $0^{\pm1}$
 & $0.247^{\pm0.007}$ & $37.6^{\pm1.1}$ & $0.37$ \\ 
                     & $3.10$
 & $1.5^{\pm0.1}$ & $0.28^{\pm0.07}$ & $0.19^{\pm0.88}$ & $3.35^{\pm1.16}$
 & $65^{\pm4}$ & $167^{\pm4}$
 & $0.253^{\pm0.018}$ & $38.5^{\pm2.8}$ & $0.17$ \\ 
Sz 129               & $0.89$
 & $179.9^{\pm0.9}$ & $0.36^{\pm0.01}$ & $-0.40^{\pm0.06}$ & $3.10^{\pm0.29}$
 & $24^{\pm1}$ & $0^{\pm1}$
 & $0.289^{\pm0.001}$ & $46.5^{\pm0.2}$ & $0.72$ \\ 
                     & $1.33$
 & $75.5^{\pm0.4}$ & $0.32^{\pm0.02}$ & $-0.19^{\pm0.08}$ & $2.34^{\pm0.21}$
 & $33^{\pm1}$ & $0^{\pm1}$
 & $0.303^{\pm0.002}$ & $48.8^{\pm0.3}$ & $0.57$ \\ 
                     & $3.10$
 & $9.5^{\pm0.2}$ & $0.15^{\pm0.10}$ & $0.11^{\pm0.52}$ & $1.18^{\pm0.43}$
 & $40^{\pm2}$ & $147^{\pm3}$
 & $0.278^{\pm0.007}$ & $44.7^{\pm1.2}$ & $0.39$ \\ 
J15592838-4021513    & $0.89$
 & $275.6^{\pm1.1}$ & $0.19^{\pm0.01}$ & $4.88^{\pm0.12}$ & $1.48^{\pm0.04}$
 & $68^{\pm1}$ & $109^{\pm1}$
 & $0.602^{\pm0.002}$ & $95.3^{\pm0.4}$ & $0.60$ \\ 
                     & $1.33$
 & $87.3^{\pm0.4}$ & $0.33^{\pm0.03}$ & $3.13^{\pm0.28}$ & $1.89^{\pm0.13}$
 & $67^{\pm1}$ & $109^{\pm1}$
 & $0.606^{\pm0.003}$ & $96.0^{\pm0.4}$ & $0.34$ \\ 
                     & $3.10$
 & $6.4^{\pm0.2}$ & $0.44^{\pm0.16}$ & $1.50^{\pm0.83}$ & $2.14^{\pm0.89}$
 & $68^{\pm1}$ & $109^{\pm1}$
 & $0.597^{\pm0.015}$ & $94.5^{\pm2.3}$ & $0.11$ \\ 
J16000236-4222145    & $0.89$
 & $122.1^{\pm1.0}$ & $0.59^{\pm0.02}$ & $-0.55^{\pm0.04}$ & $2.60^{\pm0.22}$
 & $58^{\pm1}$ & $0^{\pm1}$
 & $0.476^{\pm0.004}$ & $77.8^{\pm0.7}$ & $0.63$ \\ 
                     & $1.33$
 & $49.5^{\pm0.4}$ & $0.62^{\pm0.02}$ & $-0.52^{\pm0.04}$ & $3.47^{\pm0.35}$
 & $66^{\pm1}$ & $161^{\pm1}$
 & $0.479^{\pm0.005}$ & $78.2^{\pm0.8}$ & $0.56$ \\ 
                     & $3.10$
 & $6.5^{\pm0.2}$ & $0.45^{\pm0.13}$ & $-0.40^{\pm0.32}$ & $2.09^{\pm0.88}$
 & $56^{\pm2}$ & $164^{\pm2}$
 & $0.410^{\pm0.011}$ & $67.0^{\pm1.8}$ & $0.26$ \\ 
J16004452-4155310    & $0.89$
 & $172.5^{\pm1.2}$ & $0.44^{\pm0.03}$ & $-0.18^{\pm0.11}$ & $2.97^{\pm0.48}$
 & $73^{\pm1}$ & $59^{\pm1}$
 & $0.377^{\pm0.003}$ & $58.8^{\pm0.4}$ & $1.16$ \\ 
                     & $1.33$
 & $66.8^{\pm0.5}$ & $0.26^{\pm0.05}$ & $0.39^{\pm0.23}$ & $1.59^{\pm0.23}$
 & $66^{\pm1}$ & $54^{\pm1}$
 & $0.375^{\pm0.003}$ & $58.5^{\pm0.5}$ & $0.61$ \\ 
                     & $3.10$
 & $8.4^{\pm0.1}$ & $0.29^{\pm0.09}$ & $0.14^{\pm0.44}$ & $1.96^{\pm0.74}$
 & $59^{\pm1}$ & $54^{\pm1}$
 & $0.329^{\pm0.007}$ & $51.3^{\pm1.1}$ & $0.32$ \\ 
J16011549-4152351    & $0.89$
 & $86.9^{\pm1.4}$ & $0.81^{\pm0.03}$ & $-0.82^{\pm0.04}$ & $2.85^{\pm0.52}$
 & $67^{\pm1}$ & $156^{\pm1}$
 & $0.576^{\pm0.011}$ & $92.2^{\pm1.8}$ & ... \\ 
                     & $1.33$
 & $23.7^{\pm0.4}$ & $0.90^{\pm0.03}$ & $-0.98^{\pm0.02}$ & $3.17^{\pm0.49}$
 & $69^{\pm1}$ & $155^{\pm1}$
 & $0.587^{\pm0.010}$ & $94.0^{\pm1.6}$ & ... \\ 
                     & $3.10$
 & $2.5^{\pm0.2}$ & $0.98^{\pm0.17}$ & $-1.11^{\pm0.10}$ & $2.80^{\pm1.32}$
 & $66^{\pm3}$ & $156^{\pm3}$
 & $0.623^{\pm0.063}$ & $99.6^{\pm10.0}$ & ... \\ 
Sz 133               & $0.89$
 & $69.7^{\pm1.1}$ & $0.43^{\pm0.08}$ & $-0.18^{\pm0.16}$ & $1.88^{\pm0.38}$
 & $78^{\pm1}$ & $127^{\pm1}$
 & $0.456^{\pm0.009}$ & $70.5^{\pm1.4}$ & $1.26$ \\ 
                     & $1.33$
 & $26.6^{\pm0.2}$ & $0.49^{\pm0.03}$ & $-0.36^{\pm0.10}$ & $3.02^{\pm0.55}$
 & $77^{\pm1}$ & $127^{\pm1}$
 & $0.403^{\pm0.006}$ & $62.3^{\pm0.9}$ & $0.84$ \\ 
                     & $3.10$
 & $3.6^{\pm0.1}$ & $0.48^{\pm0.06}$ & $-0.58^{\pm0.17}$ & $3.41^{\pm1.15}$
 & $75^{\pm1}$ & $127^{\pm1}$
 & $0.365^{\pm0.011}$ & $56.4^{\pm1.7}$ & $0.44$ \\ 
J16070854-3914075    & $0.89$
 & $90.9^{\pm1.6}$ & $0.08^{\pm0.03}$ & $1.69^{\pm0.29}$ & $0.85^{\pm0.09}$
 & $59^{\pm1}$ & $0^{\pm1}$
 & $0.555^{\pm0.014}$ & $98.1^{\pm2.4}$ & $0.45$ \\ 
                     & $1.33$
 & $35.7^{\pm0.3}$ & $0.11^{\pm0.05}$ & $1.48^{\pm0.37}$ & $0.91^{\pm0.13}$
 & $59^{\pm1}$ & $0^{\pm1}$
 & $0.541^{\pm0.005}$ & $95.7^{\pm1.0}$ & $0.30$ \\ 
                     & $3.10$
 & $4.9^{\pm0.1}$ & $0.24^{\pm0.09}$ & $2.39^{\pm0.67}$ & $1.46^{\pm0.31}$
 & $71^{\pm1}$ & $156^{\pm1}$
 & $0.555^{\pm0.009}$ & $98.2^{\pm1.6}$ & $0.22$ \\ 
Sz 98                & $0.89$
 & $258.2^{\pm2.0}$ & $1.07^{\pm0.01}$ & $-0.80^{\pm0.01}$ & $4.86^{\pm0.15}$
 & $47^{\pm1}$ & $109^{\pm1}$
 & $0.730^{\pm0.006}$ & $113.6^{\pm0.9}$ & $0.26$ \\ 
                     & $1.33$
 & $107.1^{\pm0.5}$ & $0.77^{\pm0.03}$ & $-0.55^{\pm0.03}$ & $1.96^{\pm0.12}$
 & $47^{\pm1}$ & $113^{\pm1}$
 & $0.690^{\pm0.005}$ & $107.3^{\pm0.8}$ & $0.21$ \\ 
                     & $3.11$
 & $12.5^{\pm0.2}$ & $0.97^{\pm0.03}$ & $-1.01^{\pm0.04}$ & $3.54^{\pm0.66}$
 & $46^{\pm1}$ & $112^{\pm2}$
 & $0.620^{\pm0.012}$ & $96.5^{\pm1.8}$ & $0.12$ \\ 
Sz 100               & $0.89$
 & $54.0^{\pm0.5}$ & $0.16^{\pm0.04}$ & $3.54^{\pm0.90}$ & $2.31^{\pm0.51}$
 & $46^{\pm1}$ & $63^{\pm1}$
 & $0.259^{\pm0.003}$ & $35.3^{\pm0.4}$ & $0.55$ \\ 
                     & $1.33$
 & $22.0^{\pm0.2}$ & $0.20^{\pm0.03}$ & $3.72^{\pm0.73}$ & $3.45^{\pm0.61}$
 & $42^{\pm1}$ & $67^{\pm2}$
 & $0.245^{\pm0.003}$ & $33.5^{\pm0.4}$ & $0.38$ \\ 
                     & $3.10$
 & $3.1^{\pm0.1}$ & $0.28^{\pm0.06}$ & $1.01^{\pm0.63}$ & $3.26^{\pm1.09}$
 & $41^{\pm3}$ & $72^{\pm4}$
 & $0.283^{\pm0.007}$ & $38.7^{\pm1.0}$ & $0.18$ \\ 
J16083070-3828268    & $0.89$
 & $134.2^{\pm0.9}$ & $0.48^{\pm0.02}$ & $4.84^{\pm0.17}$ & $4.26^{\pm0.35}$
 & $73^{\pm1}$ & $107^{\pm1}$
 & $0.558^{\pm0.003}$ & $86.7^{\pm0.5}$ & $0.41$ \\ 
                     & $1.33$
 & $38.3^{\pm0.2}$ & $0.48^{\pm0.01}$ & $4.91^{\pm0.10}$ & $4.82^{\pm0.18}$
 & $72^{\pm1}$ & $108^{\pm1}$
 & $0.539^{\pm0.002}$ & $83.8^{\pm0.3}$ & $0.22$ \\ 
                     & $3.10$
 & $4.5^{\pm0.1}$ & $0.39^{\pm0.01}$ & $-4.88^{\pm0.14}$ & $-4.61^{\pm0.42}$
 & $72^{\pm1}$ & $107^{\pm1}$
 & $0.592^{\pm0.014}$ & $92.0^{\pm2.2}$ & $0.10$ \\ 
Sz 111               & $0.89$
 & $177.9^{\pm0.8}$ & $0.28^{\pm0.01}$ & $4.90^{\pm0.11}$ & $2.46^{\pm0.08}$
 & $54^{\pm1}$ & $44^{\pm1}$
 & $0.448^{\pm0.002}$ & $70.6^{\pm0.3}$ & $0.84$ \\ 
                     & $1.33$
 & $58.6^{\pm0.3}$ & $0.29^{\pm0.02}$ & $4.82^{\pm0.20}$ & $2.54^{\pm0.14}$
 & $54^{\pm1}$ & $44^{\pm1}$
 & $0.454^{\pm0.002}$ & $71.6^{\pm0.3}$ & $0.45$ \\ 
                     & $3.10$
 & $5.8^{\pm0.1}$ & $0.46^{\pm0.07}$ & $1.69^{\pm0.60}$ & $3.55^{\pm0.96}$
 & $52^{\pm2}$ & $41^{\pm2}$
 & $0.480^{\pm0.012}$ & $75.7^{\pm1.9}$ & $0.15$ \\ 
Sz 114               & $0.89$
 & $99.3^{\pm0.5}$ & $0.38^{\pm0.01}$ & $-0.96^{\pm0.02}$ & $4.55^{\pm0.39}$
 & $15^{\pm3}$ & $158^{\pm9}$
 & $0.247^{\pm0.002}$ & $39.9^{\pm0.4}$ & $0.60$ \\ 
                     & $1.33$
 & $42.6^{\pm0.3}$ & $0.34^{\pm0.02}$ & $-0.78^{\pm0.07}$ & $2.90^{\pm0.47}$
 & $30^{\pm2}$ & $0^{\pm1}$
 & $0.247^{\pm0.003}$ & $39.9^{\pm0.5}$ & $0.52$ \\ 
                     & $3.10$
 & $5.3^{\pm0.1}$ & $0.37^{\pm0.06}$ & $-1.04^{\pm0.16}$ & $3.02^{\pm1.28}$
 & $11^{\pm8}$ & $82^{\pm64}$
 & $0.235^{\pm0.010}$ & $37.9^{\pm1.7}$ & $0.26$ \\ 
Sz 118               & $0.89$
 & $60.8^{\pm0.6}$ & $0.32^{\pm0.05}$ & $3.23^{\pm0.80}$ & $3.61^{\pm0.90}$
 & $66^{\pm1}$ & $174^{\pm1}$
 & $0.368^{\pm0.005}$ & $60.1^{\pm0.9}$ & $0.34$ \\ 
                     & $1.33$
 & $23.6^{\pm9.6}$ & $0.33^{\pm0.38}$ & $3.09^{\pm1.18}$ & $3.67^{\pm0.81}$
 & $65^{\pm5}$ & $0^{\pm27}$
 & $0.365^{\pm0.356}$ & $59.6^{\pm58.1}$ & $0.23$ \\ 
                     & $3.10$
 & $2.9^{\pm0.1}$ & $0.28^{\pm0.08}$ & $2.34^{\pm0.87}$ & $2.33^{\pm0.88}$
 & $65^{\pm1}$ & $172^{\pm1}$
 & $0.395^{\pm0.012}$ & $64.4^{\pm1.9}$ & $0.11$ \\ 
Sz 123               & $0.89$
 & $41.3^{\pm0.5}$ & $0.22^{\pm0.02}$ & $3.13^{\pm0.63}$ & $4.10^{\pm0.65}$
 & $39^{\pm1}$ & $0^{\pm1}$
 & $0.237^{\pm0.003}$ & $38.4^{\pm0.4}$ & $0.40$ \\ 
                     & $1.33$
 & $16.2^{\pm0.2}$ & $0.24^{\pm0.01}$ & $2.27^{\pm0.45}$ & $4.42^{\pm0.57}$
 & $42^{\pm1}$ & $0^{\pm1}$
 & $0.242^{\pm0.003}$ & $39.2^{\pm0.5}$ & $0.28$ \\ 
                     & $3.08$
 & $2.4^{\pm0.1}$ & $0.21^{\pm0.01}$ & $3.95^{\pm0.57}$ & $4.05^{\pm0.43}$
 & $43^{\pm2}$ & $0^{\pm1}$
 & $0.236^{\pm0.004}$ & $38.2^{\pm0.7}$ & $0.18$ \\  \bottomrule
 \end{tabular}
 \begin{flushleft}
 \textbf{Note}
 The parameter values are the medians of their posterior distributions, and the uncertainties represent the 68\% confidence interval. 
 The $F_\nu$ posterior does not include systematic flux calibration uncertainty. 
 Note that the parameters to the right of the vertical bar ($\rhose$, $\Rse$, $\mathcal{F}$) are not free parameters in the fit, but they are rather inferred from the joint posterior on $\left\{F_\nu,\ \rho_c,\ \gamma_1,\ \gamma_2\right\}$. Detailed fit results are presented in Appendix~\ref{app:detailed.fit.results.selfsimilar}.
 A machine-readable version of this table is available online (see the Data Availability statement).
 \end{flushleft}
\label{tb:fit.results.ss2s}
\end{table*}

\begin{table*}
 \caption{Parameters of the inferred brightness profiles: Gaussian brightness.}
 \centering
 \begin{tabular}{lccccc|ccc}
   \midrule
   \toprule
Name & $\lambda$
 & $F_\mathrm{\nu}$ & $\sigma$ & $i$ & $P.A.$
 & $\rho_{68}$ & $R_{68}$ & $\mathcal{F}$\\ 
 & (mm)
 & (mJy) & ($\arcsec$) & $(^\circ)$ & $(^\circ)$
 & $(\arcsec)$ & (au) & \\ 
   \midrule
J15450887-3417333    & $0.89$
 & $42.9^{\pm0.6}$ & $0.08^{\pm0.01}$ & $35^{\pm5}$ & $64^{\pm8}$
 & $0.122^{\pm0.005}$ & $18.9^{\pm0.8}$ & $1.14$ \\ 
                     & $1.33$
 & $19.8^{\pm0.2}$ & $0.10^{\pm0.01}$ & $59^{\pm1}$ & $174^{\pm2}$
 & $0.149^{\pm0.003}$ & $23.1^{\pm0.4}$ & $1.16$ \\ 
                     & $3.10$
 & $3.7^{\pm0.1}$ & $0.07^{\pm0.01}$ & $42^{\pm10}$ & $160^{\pm15}$
 & $0.110^{\pm0.012}$ & $16.9^{\pm1.9}$ & $1.01$ \\ 
Sz 69                & $0.89$
 & $14.8^{\pm5.1}$ & $0.04^{\pm0.01}$ & $68^{\pm16}$ & $165^{\pm9}$
 & $0.068^{\pm0.006}$ & $10.4^{\pm1.0}$ & $1.61$ \\ 
                     & $1.33$
 & $7.6^{\pm0.2}$ & $0.05^{\pm0.01}$ & $51^{\pm9}$ & $134^{\pm10}$
 & $0.075^{\pm0.007}$ & $11.5^{\pm1.0}$ & $0.83$ \\ 
                     & $3.08$
 & $1.5^{\pm0.1}$ & $0.03^{\pm0.01}$ & $47^{\pm22}$ & $96^{\pm26}$
 & $0.050^{\pm0.009}$ & $7.6^{\pm1.4}$ & $1.27$ \\ 
Sz 72                & $0.89$
 & $13.9^{\pm0.2}$ & $0.05^{\pm0.01}$ & $46^{\pm16}$ & $34^{\pm16}$
 & $0.083^{\pm0.008}$ & $12.8^{\pm1.2}$ & $0.42$ \\ 
                     & $1.33$
 & $5.7^{\pm0.1}$ & $0.05^{\pm0.01}$ & $24^{\pm15}$ & $140^{\pm74}$
 & $0.081^{\pm0.005}$ & $12.6^{\pm0.8}$ & $0.27$ \\ 
                     & $3.08$
 & $0.9^{\pm0.1}$ & $0.04^{\pm0.01}$ & $49^{\pm30}$ & $46^{\pm42}$
 & $0.063^{\pm0.014}$ & $9.8^{\pm2.1}$ & $0.40$ \\ 
Sz 90                & $0.89$
 & $21.8^{\pm0.4}$ & $0.08^{\pm0.01}$ & $59^{\pm6}$ & $124^{\pm5}$
 & $0.125^{\pm0.006}$ & $19.9^{\pm1.0}$ & $0.46$ \\ 
                     & $1.33$
 & $8.7^{\pm0.1}$ & $0.10^{\pm0.01}$ & $57^{\pm3}$ & $141^{\pm3}$
 & $0.146^{\pm0.005}$ & $23.4^{\pm0.7}$ & $0.27$ \\ 
                     & $3.10$
 & $1.1^{\pm0.1}$ & $0.09^{\pm0.01}$ & $55^{\pm11}$ & $140^{\pm10}$
 & $0.140^{\pm0.021}$ & $22.3^{\pm3.3}$ & $0.16$ \\ 
Sz 108B              & $0.89$
 & $26.9^{\pm0.6}$ & $0.10^{\pm0.01}$ & $46^{\pm5}$ & $151^{\pm6}$
 & $0.151^{\pm0.006}$ & $25.5^{\pm1.0}$ & $0.55$ \\ 
                     & $1.33$
 & $11.9^{\pm0.2}$ & $0.10^{\pm0.01}$ & $48^{\pm2}$ & $143^{\pm3}$
 & $0.154^{\pm0.004}$ & $25.9^{\pm0.7}$ & $0.45$ \\ 
                     & $3.10$
 & $1.6^{\pm0.1}$ & $0.10^{\pm0.01}$ & $53^{\pm7}$ & $4^{\pm4}$
 & $0.154^{\pm0.012}$ & $25.9^{\pm2.0}$ & $0.29$ \\ 
J16085324-3914401    & $0.89$
 & $19.9^{\pm0.4}$ & $0.06^{\pm0.01}$ & $50^{\pm8}$ & $99^{\pm11}$
 & $0.083^{\pm0.009}$ & $13.8^{\pm1.5}$ & $0.75$ \\ 
                     & $1.33$
 & $7.6^{\pm0.1}$ & $0.07^{\pm0.01}$ & $54^{\pm5}$ & $131^{\pm5}$
 & $0.101^{\pm0.005}$ & $16.8^{\pm0.9}$ & $0.47$ \\ 
                     & $3.10$
 & $1.4^{\pm0.1}$ & $0.06^{\pm0.01}$ & $29^{\pm20}$ & $96^{\pm38}$
 & $0.089^{\pm0.010}$ & $14.9^{\pm1.7}$ & $0.32$ \\ 
J16124373-3815031    & $0.89$
 & $29.6^{\pm0.5}$ & $0.07^{\pm0.01}$ & $43^{\pm7}$ & $22^{\pm10}$
 & $0.105^{\pm0.006}$ & $16.7^{\pm1.0}$ & $0.56$ \\ 
                     & $1.33$
 & $11.5^{\pm0.1}$ & $0.09^{\pm0.01}$ & $51^{\pm2}$ & $179^{\pm1}$
 & $0.131^{\pm0.003}$ & $20.8^{\pm0.5}$ & $0.37$ \\ 
                     & $3.10$
 & $1.8^{\pm0.1}$ & $0.07^{\pm0.01}$ & $55^{\pm14}$ & $20^{\pm83}$
 & $0.104^{\pm0.016}$ & $16.5^{\pm2.5}$ & $0.41$ \\    \bottomrule
 \end{tabular}
 \begin{flushleft}
 \textbf{Note}
 The parameter values are the medians of their posterior distributions, and the uncertainties represent the 68\% confidence interval. 
 The $F_\nu$ posterior does not include systematic flux calibration uncertainty. Note that the parameters to the right of the vertical bar ($\rhose$, $\Rse$, $\mathcal{F}$) are not free parameters in the fit, but they are rather inferred from the joint posterior on $\left\{F_\nu,\ \sigma\right\}$. 
 Detailed fit results are presented in Appendix~\ref{app:detailed.fit.results.gaussian}.
 A machine-readable version of this table is available online (see the Data Availability statement).
 \end{flushleft}
\label{tb:fit.results.gaussian}
\end{table*}

\begin{figure}
\centering
\includegraphics[scale=1]{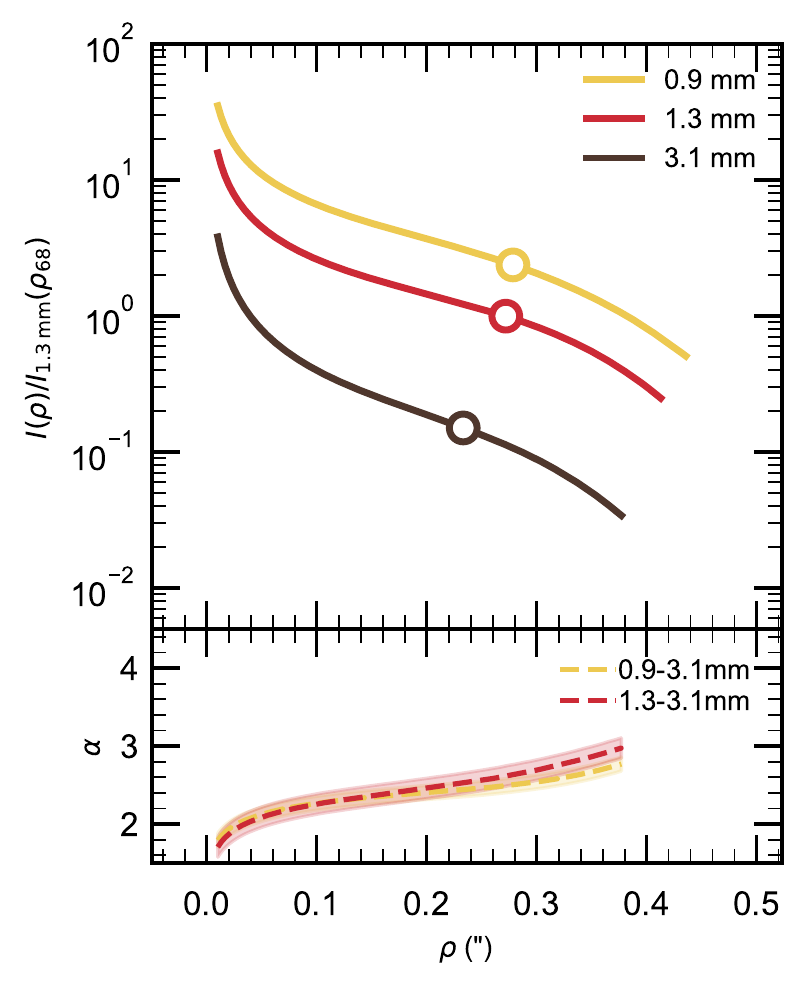}
\caption{ 
\textit{(Top)}
Brightness profiles of Sz 83 at 0.9\,mm (yellow), 1.3\,mm (red), and 3.1\,mm (black).
The profiles are normalised to the brightness measured at 1.3\,mm at the radius enclosing 68\% of the 1.3~mm emission: $1.18\times10^{10}$ Jy sr$^{-1}$. The circles represent $\Rse$.
\textit{(Bottom)}
Spectral index profile between 3.1\,mm and 0.9\,mm (yellow dashed) and between 3.1\,mm and 1.3\,mm (red dashed).
}
\label{fig:fit.brightness.images.Sz_83.maintext}
\end{figure}

As an example, here we present the results obtained for Sz~83, and we collate the results for all the other discs in Appendix~\ref{app:detailed.fit.results}. For each disc, the fits at 0.9, 1.3, and 3.1\,mm are performed independently. For each wavelength, once the MCMC has converged, we build the posterior distribution of the brightness profile by drawing 200 random samples from the posterior of the free parameters. The top panel in \figref{fig:fit.brightness.images.Sz_83.maintext} shows the median brightness profiles that we obtain for Sz~83. To ease the comparison among the profiles at multiple wavelengths, we normalise them to $I_\mathrm{1.3mm}(\rhose)$, i.e. the brightness inferred at 1.3\,mm at the location of the 1.3\,mm effective radius. The location of $\rhose$ at each wavelength is marked as an empty circle on the brightness profiles, which are plotted for $\rho\leq\rho_{95}$. The bottom panel presents the radial profile of the spectral index between 0.9 and 3.1\,mm and between 1.3 and 3.1\,mm, derived from the median profiles shown in the top panel.  

We note that these ALMA interferometric observations are affected by an absolute flux calibration uncertainty of $5\%$ at Band~3 \citep{Tazzari:2020aa}, $10\%$ at Band~6 \citep{Ansdell:2018aa}, and $10\%$ for Band~7 \citep{Ansdell:2016qf}. This reflects in a systematic uncertainty on the spectral index profiles of about $\pm$0.09 for 0.9-3.1\,mm and $\pm$0.13 for 1.3-3.1\,mm. The shaded areas around the median spectral index profiles visualise this systematic uncertainty.

\begin{figure}
\centering
\includegraphics[scale=1]{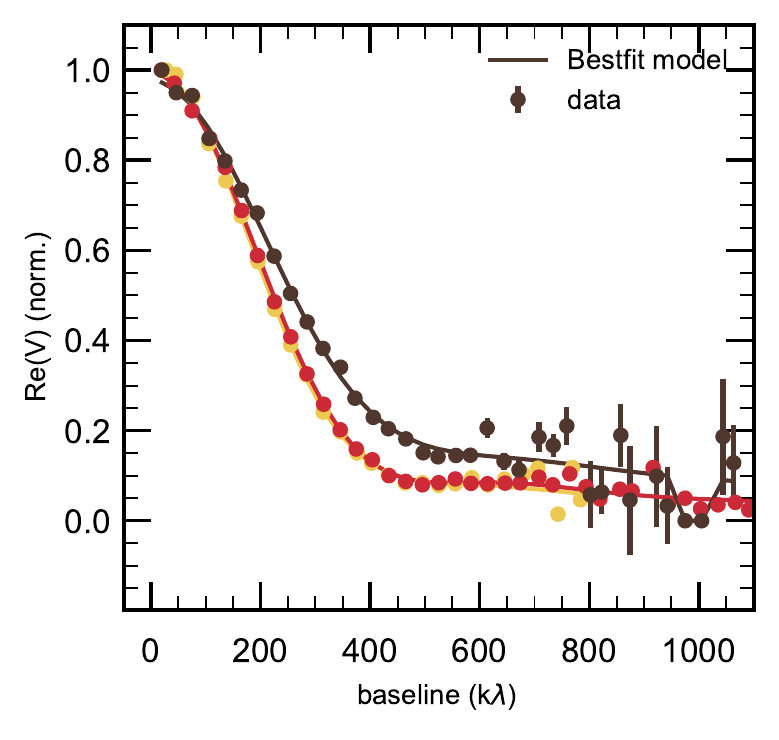}
\caption{
Comparison between the observed and bestfit model visibilities (Real part) for Sz 83 as a function of deprojected baseline at 0.9, 1.3, and 3.1\,mm. 
Data (filled circles) and model visibilities (solid lines) have been deprojected using the inferred $i$, $PA$ at each wavelength and binned in 30k$\lambda$ intervals. Colours are the same as in \figref{fig:fit.brightness.images.Sz_83.maintext}.
}
\label{fig:fits.uvplot.Sz_83.maintext}
\end{figure}

\figref{fig:fits.uvplot.Sz_83.maintext} shows the comparison between the observed visibilities and those computed for the bestfit model (i.e., the median brightness). For each wavelength, the visibilities are normalised to the disc integrated flux $F_\nu$, binned in 30k$\lambda$ intervals, and deprojected according to the inferred $i$, $PA$ values. This plot provides a useful benchmark on the quality of the fits as it compares directly the fitted data with the inferred model. In the case of Sz~83, as for most of the discs in the sample, we obtain an excellent agreement between the model and the data. In many cases our results highlight a striking similarity between the profiles at different wavelengths.

\begin{figure}
\centering
\includegraphics[scale=1]{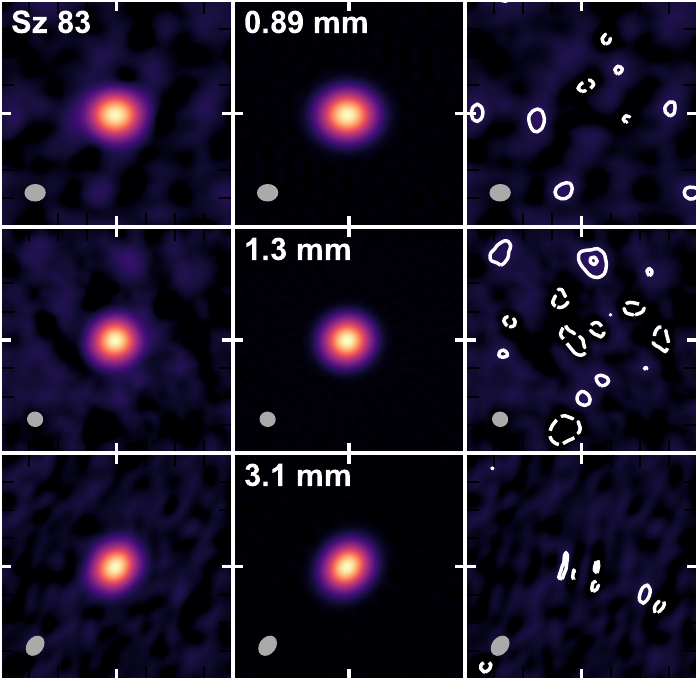}
\caption{
Synthesized images of the observed (left column), bestfit model (middle column) and residual (right column) visibilities at 0.9\,mm (top row), 1.3\,mm (middle row), 3.1\,mm (bottom row).
On each row, the color scale is normalised between the rms noise and the peak brightness of the observations image. 
Contours in the residuals image are drawn at -3, 3, 6, 12, 24, etc. times the rms noise. 
The synthesized beam is shown as a grey ellipse in each panel.
}
\label{fig:fits.images.Sz_83.maintext}
\end{figure}

As a further check on the goodness of the fits, in \figref{fig:fits.images.Sz_83.maintext} we present the synthesized images of the observed, bestfit model, and residual visibilities at 0.9\,mm, 1.3\,mm, and 3.1\,mm. The images have been produced with the \texttt{tclean} CASA command using Briggs weighting and robust parameter 0.5. In the case of Sz~83, the excellent agreement of the visibility profiles manifests in negligible (<3$\sigma$) residuals at all the wavelengths.

\section{Demographic properties}
\label{sect:results.demographic.properties}
\subsection{Disc radii: dependence on frequency}
\begin{figure}
\centering
\includegraphics[scale=1]{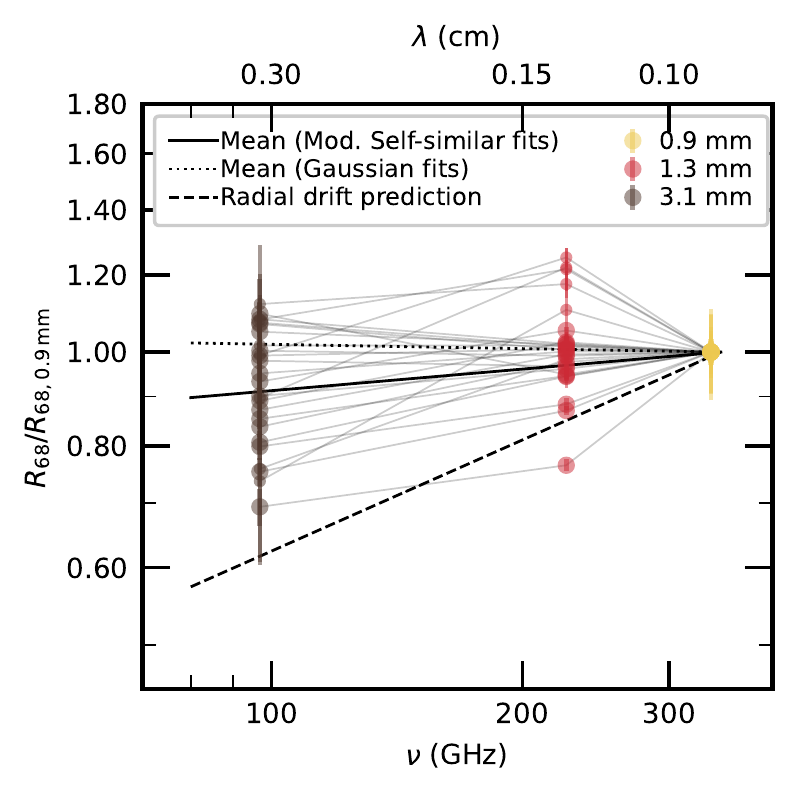}
\caption{
Disc effective radius as a function of observing frequency, normalised to the value at 0.9\,mm. The discs fitted with modified self-similar profile (Table~\ref{tb:fit.results.ss2s}) are shown as large filled circles. The discs fitted with Gaussian profile (Table~\ref{tb:fit.results.gaussian}) are shown as small filled circles. Measurements for the same disc are connected with a narrow gray line. 
The solid line represents the mean slope for the discs fitted with modified self-similar profile (\tbref{tb:fit.results.ss2s}), while the dotted line the mean slope for the Gaussian fits (\tbref{tb:fit.results.gaussian}).}
The dashed line represents the radial drift prediction by \citet{Rosotti:2019ab} discussed in Sect.~\ref{sect:implications.radial.drift}. 

\label{fig:R68_vs_freq_sample_clean_normB7_with_median_slopes}
\end{figure}

For each disc, we derive the effective radius at 0.9, 1.3, and 3.1\,mm from the inferred brightness posterior.
\figref{fig:R68_vs_freq_sample_clean_normB7_with_median_slopes} compares the multi-wavelength effective disc radii as a function of frequency. To better visualise changes across wavelengths, the radii are normalised to their values at 0.9\,mm. The most striking finding is that the majority of discs have a very similar radius across the 0.9-3.1\,mm wavelength range. Out of 26 discs, 21 of them have a 3.1\,mm radius that differs less than 20\% from their 0.9\,mm radius. Notably, the large Sz~82/IM~Lup disc is the one with the largest difference: its 3.1\,mm radius is 70\% the 0.9\,mm radius.
\tbref{tb:demographic.properties} summarises the statistics on the multi-wavelength radii. Compared to the radius measured at 0.9\,mm, the mean 1.3\,mm radius is 4\% smaller and the mean 3.1\,mm radius $\sim$9\% smaller, with both measurements being compatible with the 0.9\,mm one within 1$\sigma$. The mean of the size-frequency slopes for the whole sample is:
\begin{equation}
   \log\left(\frac{\Rse}{\mathrm{au}}\right) = \mathrm{const.} + ({0.05}\pm{0.03}) \log\left(\frac{\nu}{\mathrm{340GHz}}\right)\,,
\end{equation}
namely compatible with a flat size-frequency relation.

We note that there is a group of discs with a measured 1.3\,mm radius that appears larger than that measured at 0.9\,mm by more than 10\%: in most cases they correspond to the Gaussian fits (represented in \figref{fig:R68_vs_freq_sample_clean_normB7_with_median_slopes} with smaller filled circles), which we employed for the discs with noisier observations. 
This group of discs has a flat size-frequency relation:
\begin{equation}
    \log\left(\frac{\Rse}{\mathrm{au}}\right) = \mathrm{const.} + {-0.01}\pm{0.06} \log\left(\frac{\nu}{\mathrm{GHz}}\right)\,.
\end{equation}
The 1.3\,mm radii measurements for these discs are compatible with the 0.9\,mm radii within 1$\sigma$ and are likely due to the fainter nature of their emission.
\begin{table}
 \caption{Summary of demographic properties: radii and optically thick fraction.}
 \centering
 \begingroup
  \renewcommand{\arraystretch}{1.3} 
  \resizebox{\hsize}{!}{%
  \begin{tabular}{lcccc}
   \toprule
$\lambda$ & $\Rse$ & $\Rse/R_\mathrm{68,0.9mm}$ & $\ff$ & $\ff/\ff_\mathrm{0.9mm}$ \\ 
(mm) & (au) &  &  &  \\ 
\midrule
0.9 mm & ${56.30}\pm{7.90}$ & ${1.00}$ & ${0.68}\pm{0.07}$ & ${1.00}$ \\ 
1.3 mm & ${54.34}\pm{6.62}$ & ${0.97}\pm{0.12}$ & ${0.48}\pm{0.05}$ & ${0.70}\pm{0.07}$ \\ 
3.1 mm & ${51.88}\pm{6.54}$ & ${0.92}\pm{0.12}$ & ${0.34}\pm{0.06}$ & ${0.49}\pm{0.08}$ \\
       \bottomrule
 \end{tabular}
 }
 \endgroup
 \begin{flushleft}
 \textbf{Note.} 
The values quoted for $\Rse$  and $\ff$ are the means of the sample; 
their uncertainties are the standard error on the mean. 
\end{flushleft}
\label{tb:demographic.properties}
\end{table}
We highlight that for many discs with high signal-to-noise observations (which typically have been fitted with the modified self-similar profile), the radius is essentially constant across wavelengths. This can be seen even before modelling, just by comparing the visibility profiles, which almost perfectly overlap in many cases. 
The mean slope only for the discs that have been fitted with the modified self-similar brightness profile (typically the brightest discs in the sample) is:
\begin{equation}
    \log\left(\frac{\Rse}{\mathrm{au}}\right) = \mathrm{const.} + {0.07}\pm{0.03} \log\left(\frac{\nu}{\mathrm{GHz}}\right)\,,
\end{equation}
which is steeper than the mean for the whole sample and but still compatible with a flat size-frequency relation.

\subsection{Constraints on the optical depth}
\label{sect:opt.depth.constraints}
To quantify how much of the disc emission can be attributed to optically thick regions it is useful to introduce a new disc-averaged quantity, the optically thick fraction $\ff$, defined as the ratio between the integrated luminosity enclosed within $\rho_x$ ($x$ being the fraction defined in Sect.~\ref{sect:measuring.disc.size}) and the luminosity that would be emitted by a completely optically thick disc with a size $\rho_x$:
\begin{equation}
    \ff  = \frac{xF_\nu}{ 2\pi \cos(i)  \int_0^{\rho_x}B_\nu(T)\rho' d\rho'} = \frac{x\Lmm}{ 2\pi \int_0^{\rho_x}B_\nu(T_\mathrm{d})\rho' d\rho'}\,
\end{equation}
where $\Lmm$ is inferred from the visibility fits and we set $x=0.68$. 
Although a larger value (e.g., $x=0.95$) would presumably yield a closer approximation of the total contribution of the optically thick emission, it would be a problematic choice for the following reasons: first, since the 95\% flux-containing radius ($R_\mathrm{95}$) is much more sensitive to the slope of the outer disc than $\Rse$ is, a constraint on $\ff$ evaluated at $R_\mathrm{95}$ would be inevitably much more uncertain.
Second, due to the different signal-to-noise level of different disc observations, we would not be able to achieve a constraint on $\ff$ at $R_\mathrm{95}$ for all the discs of the sample, and for different discs we would have constraints at different $x$ values, making the set of measurements inhomogeneous.
For the dust temperature $T_\mathrm{d}$
we use the empirical parameterisation by \citet{Andrews:2013qy}:
\begin{equation}
\label{eq:tdust.andrews}
    T_\mathrm{d} = 
    T_0
    \left(\frac{L_\star}{L_\odot}\right) ^{0.25}
    \left(\frac{R}{R_0}\right)^{-q} \,,
\end{equation}
where the actual values $q=0.5$, $T_0=30\,$K $R_0=10\,$au were recently calibrated by \citet{Andrews:2018aa} using ALMA and SMA observations of discs in the Lupus and Taurus region. The stellar luminosities used for the Lupus sources are in \tbref{tb:YSOs}. We ensure that the dust temperature does not reach unrealistically low values below the threshold of $T_\mathrm{floor}=7\,$K induced by the typical interstellar radiation field in low mas star forming regions by using an \textit{effective} dust temperature equal to $T^4=T_\mathrm{d}^4+T_\mathrm{floor}^4$.
Note that $\ff$ should not be regarded as the average optical depth of the disc because, by construction, it lies between zero and one. Nevertheless it is a measure of the dominance or otherwise of optically thick emission in the integrated flux.

We highlight that $\ff$ is not directly measurable from the observations, as it requires knowledge of the \textit{size} of the disc (which we obtained through the visibility fits) and the assumption of a dust temperature profile. Compared to a simple measurement of the integrated flux ($F_\nu$), $\ff$ is intrinsically more model-dependent. However, it is a useful observational quantity that can be determined robustly from spatially resolved observations and with reasonable assumptions on the dust temperature
and, advantageously compared to $F_\nu$, leverages the information on the spatial distribution of the disc brightness.
The lack of spatial resolution that affected sub-mm/mm observations until recent years made it rarely possible to characterise discs through $\ff$. 
Here, aiming to take full advantage of the resolving power of these ALMA observations, we will use $\ff$ to gain insight into the structure of discs.

\begin{figure*}
\centering
\includegraphics[scale=1]{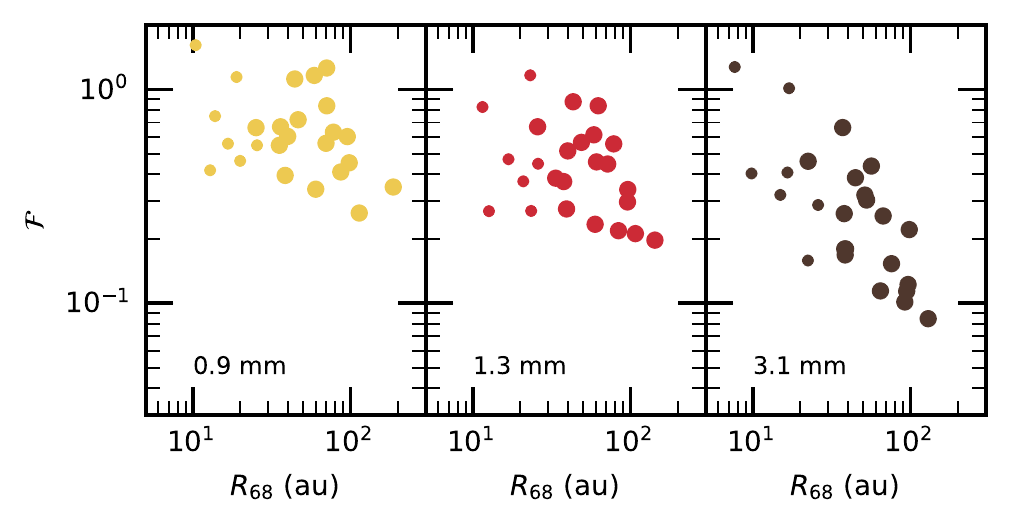}
\includegraphics[scale=1]{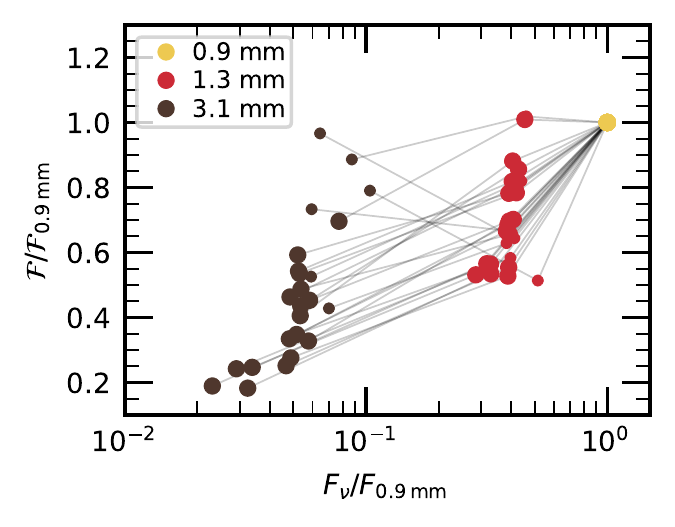}
\caption{
\textit{(Left)}
Optically thick fraction as a function of effective disc size $\Rse$ at 0.9\,mm (left panel), 1.3\,mm (center panel), and 3.1\,mm (right panel).
\textit{(Right)}
Optically thick fraction a function of integrated flux, both values normalised at 0.9\,mm. Measurements for the same disc are connected with a narrow gray line. The discs fitted with Gaussian profile (\tbref{tb:fit.results.gaussian}) are shown as small filled circles: in three of them (Sz~69, Sz~72, and J16124373-3815031) the optically thick fraction is more uncertain owing to their very compact nature and appear as outliers in the plot with $\mathcal{F}_\mathrm{3.1\,mm}\geq \mathcal{F}_\mathrm{1.3\,mm}$.}
\label{fig:ff_vs_Ftot_sample_clean}
\end{figure*}

Figure~\ref{fig:ff_vs_Ftot_sample_clean} summarises the distribution of optical depth fractions that we derive at the three wavelengths. The left panel shows, for each wavelength, $\ff$ as a function of disc effective size. As expected, the contribution of optically thick emission to the total integrated flux decreases significantly at longer wavelengths, with median $\ff$ values decreasing from $0.71 \pm 0.06$ at 0.9\,mm, to $0.49 \pm 0.05$ at 1.3\,mm, to $0.34 \pm 0.06$ at 3.1\,mm. Moreover, at 3.1\,mm we notice that there is a marked correlation for which the largest discs are also those with lowest $\ff$. The right panel shows, for each disc, $\ff$ as a function of integrated flux at the three wavelengths, with both quantities normalised to their values at 0.9\,mm. To ease the interpretation of the plot, measurements belonging to the same disc are connected with a thin grey line. The plot shows that there is a drop in optical depth at longer wavelengths.

\subsection{Millimeter continuum size-luminosity relation}
\label{sect:size.luminosity.relations}
A correlation between the millimeter continuum disc sizes ($\Rse$) and their flux at 0.9\,mm was found by \citet{Tripathi:2017aa} using SMA observations of a sample of bright discs in the Taurus region, and was later confirmed by \citet{Andrews:2018ab} using a larger complete sample of discs in the Lupus region. 
The correlation that they found, which is 
\begin{equation}
    \log\left( \frac{R_{68}}{\mathrm{au}} \right) = (2.15\pm 0.10) + (0.51\pm 0.06) \log \left[ F_\nu \left(\frac{d}{140\,\mathrm{pc}}\right)^2 \right]
\end{equation}
for the discs in the Lupus region, was interpreted as a constant surface brightness (i.e., $F_\nu \propto R^2$) with an average optically thick fraction of about 0.3. 
Here we revisit the size-luminosity correlation in the context of the multi-wavelength observations that we obtained at 0.9, 1.3 and 3.1\,mm, looking for the presence of the same scaling relation at 1.3 and 3.1\,mm. 

\begin{figure*}
\centering
\includegraphics[scale=1]{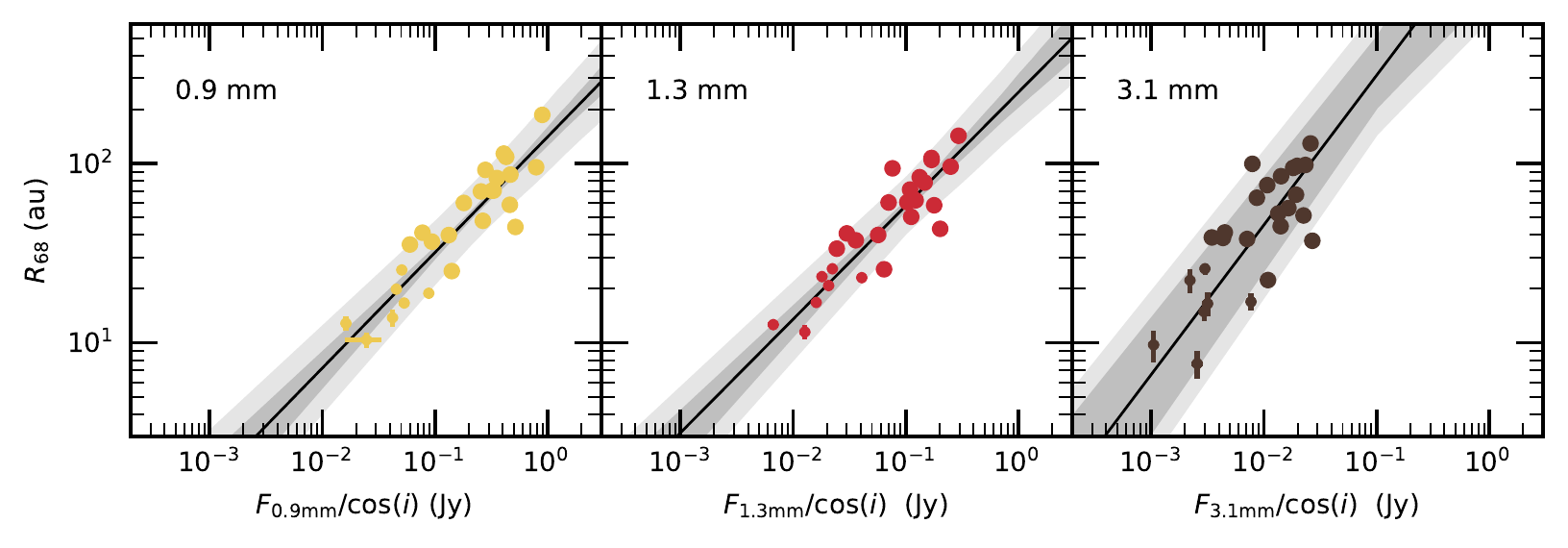}
\caption{Millimeter continuum size - luminosity relation at multiple wavelengths: 0.9\,mm (left), 1.3\,mm (center), 3.1\,mm (right). The solid black line is the median scaling relation from the Bayesian linear regression. The dark gray area represents the 68\% confidence interval around the median relation, and the light gray area includes the inferred scatter.}
\label{fig:size-luminosity.three.bands}
\end{figure*}

\figref{fig:size-luminosity.three.bands} presents the Lupus discs radii ($\Rse$) against their millimeter luminosity (re-scaled at the common distance of 150\,pc) measured at 0.9, 1.3, and 3.1\,mm. 
To quantitatively characterise the properties of the size-luminosity scaling relation, in the simple assumption of a linear correlation in the logarithmic space (i.e., a power-law correlation in the linear space), we parametrise the relation as 
\begin{equation}
\label{eq:linear.regression.def}
    \log \left(\frac{\Rse}{\mathrm{au}}\right) = \mathcal{A} +  \mathcal{B} \log\left[ \frac{F_\nu}{\cos i}\left(\frac{d}{150\,\mathrm{pc}}\right)^2 \right] + \epsilon
\end{equation}
where $\mathcal{A}$ is the intercept (normalisation), $\mathcal{B}$ is the slope (power-law index) of the relation, and $\epsilon$ is a Gaussian scatter term along the y axis ($\Rse$, in this case).
The implications of the $\cos(i)$ term are discussed at the end of this Section.
We perform a Bayesian linear regression with a mixture of 2 Gaussian generative models following the method by \citet{Kelly:2007lq} and using the implementation in the 
\texttt{linmix}
\footnote{Code available at \href{https://github.com/jmeyers314/linmix}{https://github.com/jmeyers314/linmix}.} 
Python package. Table~\ref{tb:linear.regressions} summarises the properties of the posterior inferred for the $\mathcal{A}$, $\mathcal{B}$ parameters, as well as for $\sigma$ (the standard deviation of the scatter term) and for the correlation coefficient $\rho$. The corner plots of the MCMC used for the linear regression show that the posteriors of these parameters are single-peaked and well behaved. 

\begin{table}
 \caption{Results of the linear regressions for the size-luminosity relation.}
 \centering
 \begingroup
  \setlength{\tabcolsep}{10pt} 
  \renewcommand{\arraystretch}{1.5} 
     \resizebox{\hsize}{!}{%
  \begin{tabular}{lcccc}
   \toprule
$\lambda$ & $\mathcal{A}$ & $\mathcal{B}$ & $\sigma$ & $\rho$ \\ 
\midrule
0.9\,mm & ${2.15}^{+0.06}_{-0.06}$ & ${0.64}^{+0.08}_{-0.07}$ & ${0.14}^{+0.03}_{-0.02}$ & ${0.91}^{+0.04}_{-0.05}$ \\ 
1.3\,mm & ${2.40}^{+0.10}_{-0.09}$ & ${0.63}^{+0.08}_{-0.08}$ & ${0.13}^{+0.03}_{-0.02}$ & ${0.91}^{+0.04}_{-0.06}$ \\ 
3.1\,mm & ${3.32}^{+0.43}_{-0.37}$ & ${0.83}^{+0.22}_{-0.19}$ & ${0.15}^{+0.05}_{-0.05}$ & ${0.88}^{+0.07}_{-0.12}$ \\ 
       \bottomrule
 \end{tabular}
 }
 \endgroup
 \begin{flushleft}
 \textbf{Note.} 
 The values quoted for $\mathcal{A}$ (intercept), $\mathcal{B}$ (slope), $\sigma$ (scatter), and $\rho$ (correlation coefficient)
 are the medians of their posterior distribution; their uncertainties are the central 68\% confidence interval.
 \end{flushleft}
\label{tb:linear.regressions}
\end{table}

The results of the linear regressions performed at 0.9, 1.3, and 3.1\,mm can be summarised as follows: 
\begin{enumerate}[(i)]
 \item 
 at 0.9\,mm: we confirm the presence of a very tight correlation (correlation coefficient $\rho=0.91\pm0.04$). Although we do not recover exactly the same slope reported  by \citet{Andrews:2018ab}, our regression is compatible within 2$\sigma$ with their results. The somewhat steeper slope that we find is likely due to the fact that we miss some of the faintest targets in the sample analysed by \citet{Andrews:2018ab}, for which they typically infer uncertain disc sizes that are compatible with very large values: overall, these faint (and possibly large) discs are likely to have a flattening effect on the size-luminosity correlation. 
 \item 
 at 1.3\,mm: we discover a correlation that is essentially identical to that at 0.9\,mm, except for a larger normalisation. 
 \item 
 at 3.1\,mm: we discover a new correlation, which is steeper than the ones at shorter wavelengths (slope is 0.83 as opposed to 0.63) and has a significantly larger normalisation. Due to the larger uncertainties of the fainter 3.1\,mm emission, the slope inferred at 3.1\,mm is still formally compatible within 2$\sigma$ with the slopes at shorter wavelengths. 
\end{enumerate}
An immediate interpretation for the increase in the intercept  from 0.9 to 3.1\,mm is that overall the emission becomes more optically thin at longer wavelengths: for a given disc size, discs emit fainter and optically thinner emission at longer wavelength. This is in line with the results discussed in the previous section and shown in \figref{fig:ff_vs_Ftot_sample_clean}. The difference in slope, instead, suggests that there is a systematic effect for which the emission of large and small discs behaves differently across wavelengths, with most of the difference occurring between 1.3 and 3.1\,mm.  

Using models of grain growth and drift, \citet{Rosotti:2019aa} found that a disc size - millimeter luminosity relation is to be expected if radial drift is the dominant mechanism setting the maximum grain size. They predict that such relation should have the same slope if observed at different wavelength: although this is what we find for the 0.9-1.3\,mm wavelength range, between 1.3 and 3.1\,mm we observe a steeper slope. 
In the same scenario, they predict that the intercept of the size-luminosity relation should scale as $\lambda^2$ ($\lambda$ being the observing wavelength). We find that there is a broad agreement with this result, with the observed intercepts being within 20\% of the expected values.

\begin{figure*}
\centering
\includegraphics[scale=1]{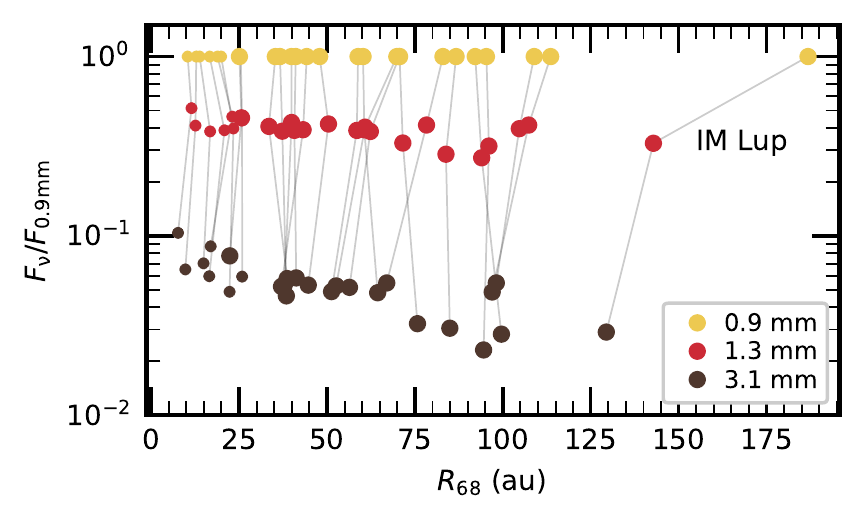}
\includegraphics[scale=1]{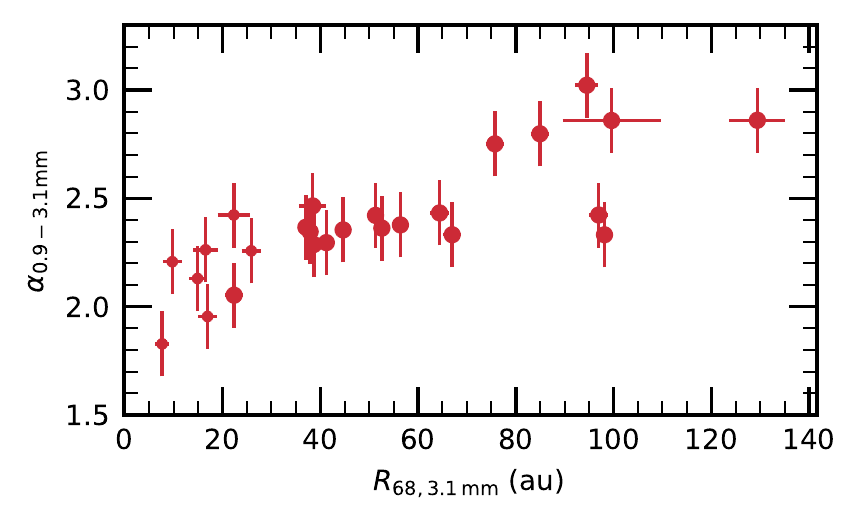}
\caption{
\textit{(Left)}
Luminosity normalised at 0.9\,mm as a function of effective disc size, at 0.9\,mm (yellow), 1.3\,mm (red), 3.1\,mm (black). Measurements for the same disc are connected with a thin solid gray line.
\textit{(Right)}
Spectral index between 0.9 and 3.1\,mm as a function of effective disc size measured at 3.1\,mm.
}
\label{fig:lum.vs.size.and.alpha.vs.size}
\end{figure*}

To further investigate the presence of a systematic effect across wavelengths, in the left panel of  \figref{fig:lum.vs.size.and.alpha.vs.size} we plot the disc luminosity at the three wavelengths (normalised to the 0.9\,mm value) as a function of the disc size measured at the same wavelength. It is evident that there is a systematic trend in which the luminosity at 3.1\,mm consistently decreases by larger amounts as the disc size increases. 
Another way to look at the same data is shown in the right panel of the same Figure, which reports the 0.9-1.3\,mm spectral index as a function of the disc size measured at 3.1\,mm. Despite two outliers around 100\,au, it is clear that the spectral indices tend to be larger for larger discs.

We shall note that, compared to \citet{Andrews:2018ab},  here we re-scale the integrated flux by $\cos(i)$: effectively, we are considering the \textit{face-on} luminosity of the discs (i.e. an intrinsic disc property) rather than their integrated flux (which inevitably includes the effects of the viewing geometry). It is worth noting that theoretically only the optically thick part of the emission scales with  $\cos(i)$, while the radiation emerging from the optically thin parts is independent of the viewing angle. 
For completeness, we also perform the linear regression \textit{without} the $\cos(i)$ correction (namely, as in \citealt{Andrews:2018ab}): the results are reported in Appendix~\ref{app:size.luminosity.inc.correction}. Interestingly enough, we find that the relations between disc size ($\Rse$) and \textit{face-on} luminosity (presented in this section, results in \tbref{tb:linear.regressions}) are much tighter (scatter $\sigma$ is halved) than the relations between their size and integrated flux (results in \tbref{tb:linear.regressions.no.inc.corr}). 
The tightening of the size-luminosity relation compared to the size-flux one suggests that the discs have a substantial optically thick contribution to their emission, broadly in line with the rather high optically thick fractions that we observe (Sect.~\ref{sect:opt.depth.constraints}). We defer the discussion of more detailed implications of this effect to dedicated future investigations.

\section{Interpretation of the data}
\label{sect:discussion}
The most striking result from our observational dataset is the fact that there is very little variation in disc size
as measured at wavelengths in the range $0.9-3.1$mm (see Figure \ref{fig:R68_vs_freq_sample_clean_normB7_with_median_slopes}).
We first consider whether this result can be  explained in terms of disc truncation and then discuss whether
it is compatible with the expectations of grain growth models and radial drift in smoothly structured discs. Having
found that neither of these scenarios are compatible with the data we
undertake a more detailed exploration of the combination of parameters that are consistent  with
the observed properties of our sample.

\subsection{Compatibility with truncated discs {models}}
\label{sect:truncated.models}
An obvious scenario that could produce a truncation in the disc brightness across multiple wavelengths would be a genuine drop in dust surface density, where $\Rse$ would reflect a \textit{physical} disc outer edge.
Our sample partially overlaps that of the high resolution DSHARP survey (7 common sources, see note 3 in Table~\ref{tb:YSOs}) and so we can check whether such a steep decline in emissivity is indeed characterising our sources. We note that our
finding that $\Rse$ is constant with wavelength is common to discs that are seen to be highly structured in
DSHARP images  (e.g., IM Lup) and also to those showing modest substructures and a smooth decline in
surface brightness in the outer disc (e.g. MY Lup, Sz 114). We thus conclude that, although a surface density truncation could explain the constancy of $\Rse$ in some discs (namely, those that exhibit a sharp continuum outer edge), some further explanation is needed to account for the constancy of $\Rse$ in the whole sample.

\subsection{Compatibility with radial drift models in smooth
discs}
\label{sect:implications.radial.drift}
Another way 
to produce an $\Rse$ that is roughly constant with wavelength
is the case in which the surface brightness profiles
at the three wavelengths are close to being
scaled versions of each other, as appears to be the case, for example, in the case of
Sz~83 (\figref{fig:fits.uvplot.Sz_83.maintext}). This is equivalent to saying that there is little radial variation in the spectral
index, as is illustrated in the lower panel of \figref{fig:fits.uvplot.Sz_83.maintext}. 

This is however
surprising in the context of models for grain growth and evolution which predict a rising spectral
index with radius, due both to optical depth effects and the evolution of the grain size distribution towards a larger maximum grain size in the inner disc. For example, \citet{Rosotti:2019ab} found that, in such models, $\Rse$ at a given
wavelength is closely related to the location in the disc at which the maximum grain size
corresponds to an opacity resonance at that wavelength (i.e. where $\amax \sim \lambda/2\pi$), since the opacity drops steeply for grain populations
with smaller values of $\amax$. This \textit{cliff} in disc opacity (of amplitude $8-10$ for compact grains)
should imprint itself on the surface brightness profile regardless of the disc optical depth (unless the disc is so dense that it is optically thick even outside the opacity cliff, a situation that is never encountered in the case of smooth disc profiles with realistic parameters). 
In the case that the maximum grain size decreases with radius, the location of the opacity cliff is expected to move inwards at longer observing wavelengths, where the grain size corresponding to the opacity feature is larger. Thus the clear implication is that discs are expected to \textit{decline} in size as the wavelength of observations changes from 0.9 to 3.1\,mm. 

In \figref{fig:R68_vs_freq_sample_clean_normB7_with_median_slopes} we plot the size-wavelength relation predicted in the radial drift regime, computed as the the average of the radial drift dominated models presented in \citet{Rosotti:2019ab}: the expectation is indeed far from the observed properties of the sample, predicting much smaller radii at 1.3 and 3.1\,mm than observed. 
For reference, with respect to $\Rse$ measured at 0.9\,mm, this would correspond to a 17\% smaller radius at 1.3\,mm and a 40\% smaller radius at 3.1\,mm.

\begin{figure}
\centering
\includegraphics[scale=1]{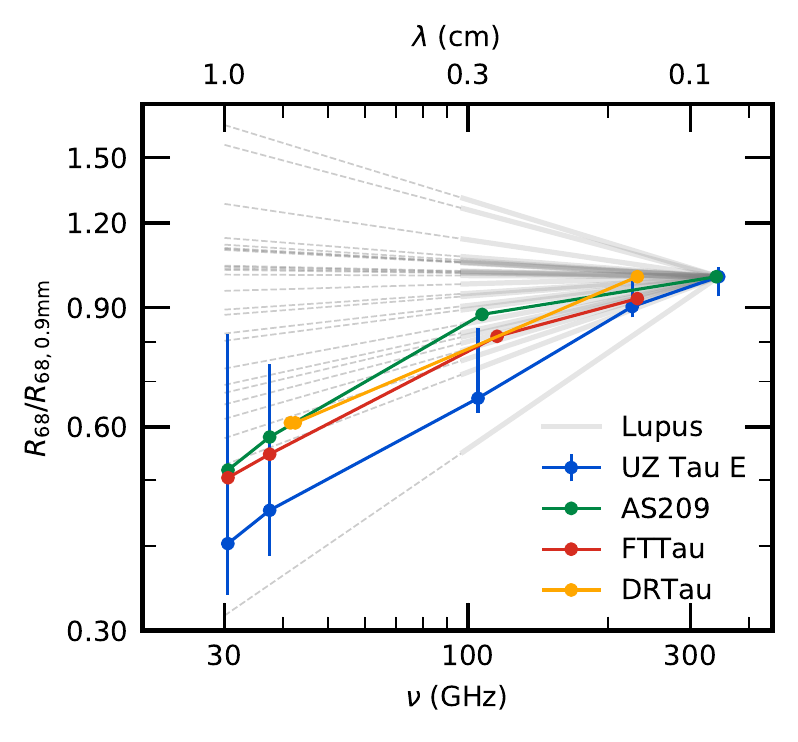}
\caption{
Comparison of the size-frequency slopes for Lupus discs (grey lines) and literature measurements (coloured lines).
Thick grey lines represent the slopes inferred for for Lupus discs in this study, with extrapolation to longer wavelengths as dashed grey lines.
}
\label{fig:size-frequency.comparison}
\end{figure}

Our Lupus dataset indicates that the grain populations probed by 0.9-3.1\,mm observations are typically co-located.
In \figref{fig:size-frequency.comparison} we depict the Lupus data (grey lines) extrapolated to longer wavelengths (dashed grey lines) and show that even at wavelengths of 1 cm, the predicted change in size is still relatively modest (less than a factor of $2$). 

In \figref{fig:size-frequency.comparison} we also show the results of previous multi-wavelength measurements of disc sizes from the literature \citep{Perez:2012fk,Tripathi:2018aa,Tazzari:2016qy}, which employed observations from the Submillimeter Array (SMA), the Combined Array for Research in Millimeter-wave Astronomy (CARMA) and the Karl G. Jansky Very Large Array (VLA), covering a wavelength range from $\sim0.88\,$mm to 1\,cm. 
It is striking that in the 0.88-3.1\,mm wavelength range, all these literature measurements are consistent within the uncertainties with the distribution of size-frequency relations inferred for the Lupus sample.
The agreement holds if we extend the comparison to 1\,cm wavelength, with the extrapolated Lupus slopes.
Note that so far only the brightest (and thus largest) discs have been probed at long ($\lambda>7$\,mm) wavelengths.

The homogeneous analysis that we present here for a sample of Lupus discs sheds new light on this issue by revealing that there is a sizeable fraction of the disc population with a significantly flat size-frequency relation, in which the bright objects probed so far at long wavelengths lie at the steeper end of the distribution.

\subsection{An exploration of viable disc parameters}
Given the difficulty of explaining the near wavelength-independence of $\Rse$ in this
dataset we now resort to performing a suite of simply parametrised models in order
to explore what emissivity properties  could in principle explain our results.
Given the limited resolution of our dataset, it is not fruitful to undertake detailed
modeling of radial profiles of spectral indices of individual sources (cf. \citealt{Carrasco-Gonzalez:2019aa}). Instead we want to consider a set
of global properties of each disc that can we can reliably extract from our observations
and then consider what sorts of models are compatible with the ensemble. In order
to assess the requirements that apply to discs with a variety of brightnesses and radial sizes we consider three \textit{dimensionless} numbers: 
\begin{enumerate}[(i)]
    \item the ratio between $\Rse$ at 3.1\,mm and $\Rse$ at 0.9\,mm;
    \item $\alphamm$, the disc-averaged spectral index;
    \item $\ff_\mathrm{1.3\,mm}$, the optically thick fraction at 1.3\,mm. 
\end{enumerate}
As in many previous studies \citep{Perez:2012fk,Tazzari:2016qy,Tripathi:2018aa,Huang:2020aa} we  can regard the spectral index as containing information both about the emissivity properties of the dust and/or the prevalence of optically thick emission, while the optically thick fraction primarily  provides information on the latter issue. Since neither of these quantities can be simply mapped onto a unique disc emissivity profile, we conduct some simple forward modelling in order to constrain the types of parameters that match the values of the above mentioned three quantities in the Lupus dataset. 
Figure \ref{fig:toy_model.smooth.grid.beta.const} presents the data in the plane of optically thick fraction at $1.3$\,mm versus spectral index between $0.9$ and $3.1$\,mm and shows that it tends to be clustered at moderate values of the optically thick fraction
and low values of spectral index, with the bulk of the population residing in the range:
\begin{equation}
\begin{split}
  0.2&\leq \ff_\mathrm{1.3\,mm} \leq 0.6\quad\textrm{and}\\ 
  2.4&\leq \alphamm \leq 3.0\,.
\end{split}
\end{equation}
 
We now run a series of simple models with different optical depth radial profiles and search for those models that best reproduce these observational constraints. Note that in the following two sections our simple models neglect the role of scattering of mm emission. Inclusion of scattering would not change the predictions of the models that we find to be viable in the next two Sections since the modest observed optically thick fractions for most sources does not {\it require} the dominance of optically thick emission (where scattering modifies the spectral index). 
Nevertheless, an alternative explanation of the available data is one where the emission is dominated by optically thick emission with low areal filling factor where the scattering albedo is sufficiently high (requiring grains larger than 1 mm). We discuss the potential effect of scattering in more detail in Appendix~\ref{app:role.of.scattering}.
 
\subsubsection{Smooth radial profiles}
\label{sect:toymodel.smooth}
We first run models with the optical depth described by a modified self similar spatial profile, and a power-law spectral dependency:
\begin{equation}
\tau_\nu = \tau_{\nu0} \left(\frac{R}{R_c}\right)^{\gamma_1} \exp\left[-\left(\frac{R}{R_c}\right)^{\gamma_2}\right]
\left(\frac{\nu}{\nu_0}\right)^{\beta(R)}\,,
\label{eqn:tauprescript}
\end{equation}
where $\beta(R)$ increases linearly from 0 in the inner disc to an asymptotic value $\beta_\mathrm{out}$ at a radius $R_\beta$, namely:
\begin{equation}
    \beta(R) = 
    \begin{cases}
        \beta_\mathrm{out} \frac{R}{R_\beta}  & R\leq R_\beta \\
        \beta_\mathrm{out}                     & R> R_\beta
    \end{cases}
\end{equation}   
This implementation gives us the flexibility to realise a simple scenario in which $\beta$ (and therefore the maximum grain size) is constant throughout the disc ($R_\beta=0$, $\beta=\beta_\mathrm{out}$), as well as the scenario in which grains are  larger in the disc interior (within $R_\beta$). 

For the dust temperature we use the parameterisation in \eqref{eq:tdust.andrews}. The brightness of the disc at a given observing frequency $\nu$ is computed as
\begin{equation}
    I_\nu = B_\nu(T)\left[1-\exp(-\tau_\nu/\cos i)\right]\,.
\end{equation}
In these toy models we assume a face on disc ($i=0$), a Solar luminosity star ($L_\star=1L_\odot$), $\gamma_1=-1$, $\gamma_2=1$, and $R_c =30\,$au (note that the model trajectories in the $\ff_\mathrm{1.3\,,mm}-\alphamm$ plane are only very weakly dependent on $R_c$ since the models are scale-free apart from the weak dependence on $R_c$ introduced by the temperature parametrisation.

For a given value of $R_\beta$, we compute a grid of models where we vary the optical depth normalisation $\tau_{\nu0}=10^{-3},10^{-2},10^{-1},1,10,100$, the outer dust spectral index value $\beta_\mathrm{out}$ from 0 to 4 in steps of 0.5. Note that we set $\nu_0=345$\,GHz, corresponding to the observing wavelength of $0.869$\,mm. 

\begin{figure}
\centering
\includegraphics[scale=1]{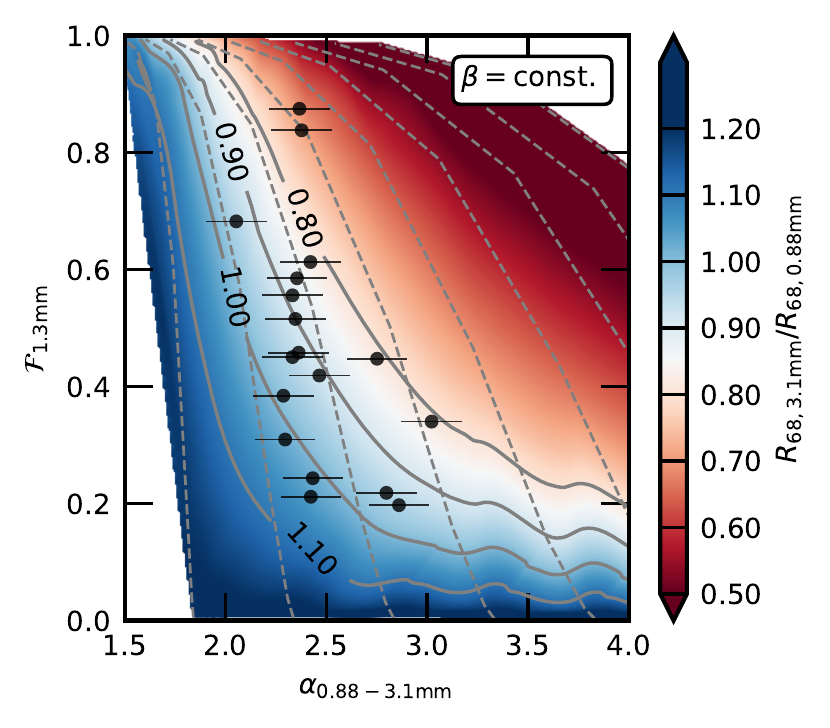}
\caption{
Optically thick fraction at 1.3\,mm as a function of 0.9-1.3\,mm spectral index for a grid of models with $\beta$ constant throughout the disc ($R_\beta=0$). Dashed gray lines connect models with same $\beta$, from $\beta=0$ (left-most) to $\beta=4$ (right-most) in steps of 0.5. The coloured map shows the ratio of disc size at 3.1 and 0.9\,mm, with labelled contours ranging from  1.1 to 0.8.
}
\label{fig:toy_model.smooth.grid.beta.const}
\end{figure}

\medskip
\noindent\textbf{Radially constant $\beta$}\quad
\figref{fig:toy_model.smooth.grid.beta.const} presents the model results for the case of $R_\beta=0$, i.e. $\beta$ is spatially constant.
The dashed lines in \figref{fig:toy_model.smooth.grid.beta.const} each represent a sequence of models with increasing optical depth normalisation at fixed $\beta_\mathrm{out}$. Each dashed line intersects the $x-$axis at the spectral index for optically thin emission with $\beta=\beta_\mathrm{out}$. The fact that this intersection occurs at a value of $\alphamm$ that is somewhat less than $2+\beta$ is a consequence of the fact that the outer parts of the disc are not entirely in the Rayleigh-Jeans regime. \figref{fig:toy_model.smooth.grid.beta.const} also depicts contours of the ratio of $\Rse$ at $3.1$ to $0.88$\,mm, which is also denoted by the colour scale and where the pale shadings correspond to a ratio near unity. 
Detailed properties for some models representative of different regimes in the $\ff_\mathrm{1.3\,mm}-\alphamm$ plane are presented in Appendix~\ref{app:detailed.toymodel.results:smooth}.
 
Figure \ref{fig:toy_model.smooth.grid.beta.const} immediately demonstrates that the models that pass through the region of the $\ff_\mathrm{1.3\,mm}-\alphamm$ plane occupied by the data also automatically satisfy the requirement of having similar $\Rse$ values in the two wavebands. Most of these successful models have input $\beta_\mathrm{out}$ values in the range $0.5-1$, with a few sources being compatible with $\beta$ up to 1.5. We have experimented with a variety of monotonically declining surface density profiles and this conclusion remains robust (see below). Such a range of acceptable $\beta$ values is unsurprising given that the observed $\alphamm$ values are rather low and yet the optically thick fraction is insufficiently high for these low $\alphamm$ values to be explicable purely in terms of high optical depth. It is also unsurprising that a constant $\beta$ model with only moderate optical depth should yield a wavelength independent disc radius since the radial profile of spectral index is in this case rather flat and hence the emissivity profiles in the two wavebands are very similar.

\begin{figure*}
\centering
\includegraphics[height=6.2cm, trim=0 0 1.8cm 0, clip]{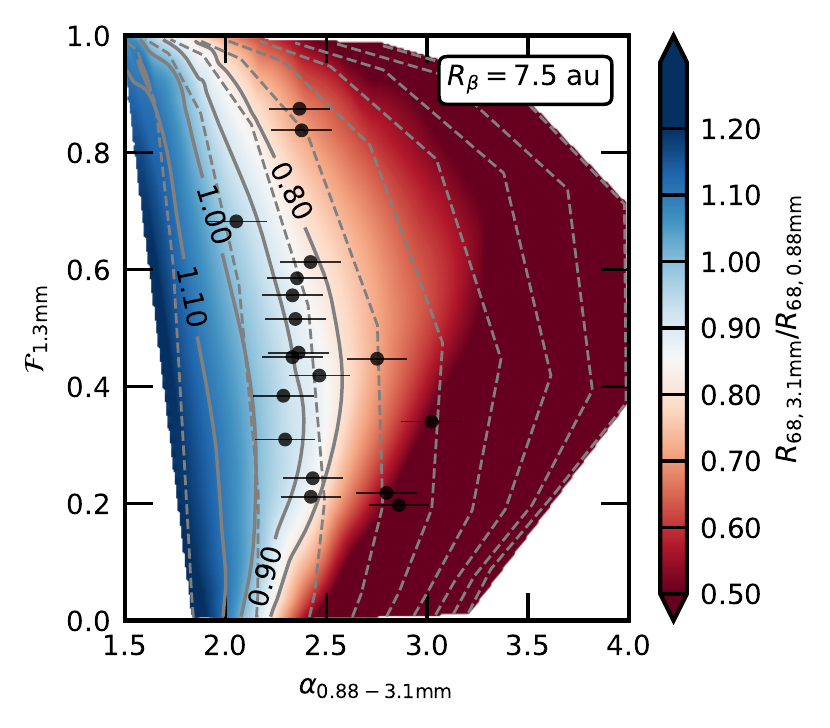}
\includegraphics[height=6.2cm, trim=0.5cm 0 1.8cm 0, clip]{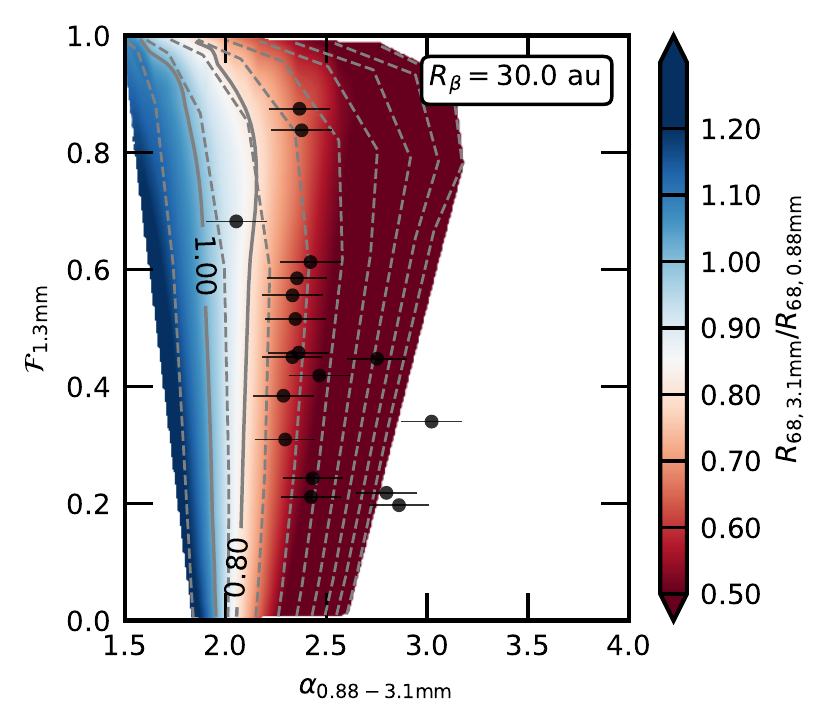}
\includegraphics[height=6.2cm, trim=0.5cm 0 0 0, clip ]{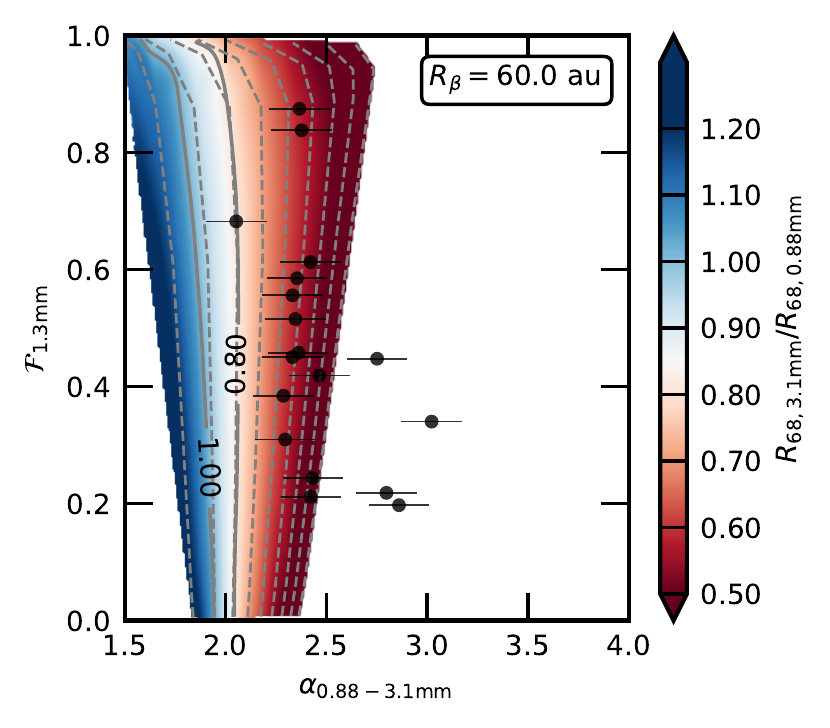}
\caption{
Model grids for different $R_\beta$ values. Colours, lines, and data points follow the same definitions as in \figref{fig:toy_model.smooth.grid.beta.const}. 
The dashed lines in each panel correspond to different values of $\beta_\mathrm{out}$, from $\beta_\mathrm{out}=0$ (left-most) to $\beta_\mathrm{out}=4$ (right-most) in steps of 0.5.
}
\label{fig:toy_model.smooth.grid.beta.linear}
\end{figure*}

\medskip
\noindent\textbf{Radially increasing $\beta$}\quad
We now relax the assumption of radially constant input $\beta$. 
Each panel in Figure \ref{fig:toy_model.smooth.grid.beta.linear}
has a fixed value of $R_{\beta}$ and the various dashed lines in each panel correspond to different values of $\beta_\mathrm{out}$. 

It is immediately obvious that large values of $R_\beta$ (as in the lower two panels) fail to reproduce the data. Firstly, they have too much emission from low $\beta$ material at small radius  to be able to replicate the larger $\alpha$ values in the sample. However, what is more restrictive is the set of constraints imposed  by the requirement that the radii at $3.1$mm
and $0.88$mm should be nearly equal. As
can be seen particularly in the lower-left hand panel, there are models that pass through the data in the $\ff_\mathrm{1.3\,mm}-\alphamm$ plane but where the radius at $3.1$\,mm is substantially smaller (factor two) than that at $0.88$\,mm. These models fail because they predict too much variation in $\beta$ over a region of the disc that is still contributing significantly to the total flux. This translates into a radially variable spectral index and the resulting difference in brightness profiles in the two bands then leads to different corresponding $\Rse$ values.
To test the robustness of this result we have computed the grid of models for a range of $\gamma_1, \gamma_2$ values, reproducing discs with significantly flatter interior ($\gamma_1=-0.3$) and (or) steeper outer edge ($\gamma_2=3$): although the actual value of $R_\beta$ that best fits the data changes slightly from case to case, the results that we have just presented do not change.

It is notable that this modeling rules out a scenario which has been proposed to explain spectral index data, where the inner disc is optically thick while the outer disc consists of a region of optically thin emission with small grains (for which $\beta =2$): see, e.g., \citet{Ricci:2012aa}. This combination can reproduce intermediate values of the spectral index as well as the observed values of the optical depth fraction, as seen, for example, in the lower left panel of Figure \ref{fig:toy_model.smooth.grid.beta.linear} where contours with $\beta_\mathrm{out}=2$ pass through the region occupied by the observational data. However, it can be seen that this region of the plot is shaded dark red, indicating that the predicted R68 values at $0.9$ and $3.1$\,mm are very different. This can be readily understood in that the flux profiles at each wavelength are shaped by the radius at which the emission makes the transition from being optically thick to thin. In the case of high $\beta$, the significantly lower opacity at $3.1$mm drives this transition to smaller radius and hence results in a steeply declining disc size as a function of observing wavelength. 
    
We therefore conclude from this simple exercise that the way to reproduce the typical spectral indices, optically thick fractions and multi-wavelength $\Rse$ ratios of the Lupus discs is to invoke a dust distribution where the value of the opacity index  $\beta$ is in the range of $0.5-1$ for most sources, at least over a substantial fraction of the disc, although some lower $\beta$ material at smaller radius is also allowed. We will discuss the implications of this result for the properties of grains in discs in Section~\ref{sect:dustprop} below but now turn to the question of how these conclusions would be modified in the case of discs with significant substructure.

\subsubsection{Structured radial profiles: the case for small grains}
\label{sect:toymodel.structured}
We have shown  that a class of models that satisfies  the observational constraints listed above is one in which $\beta$ is in the range $0.5-1$ over much of the disc, translating into a spectral index profile that plateaus with values in the range $\sim 2.5-3$. If this is interpreted in terms of grain properties we will see in Section~\ref{sect:dustprop} that this corresponds to \textit{large} grains ($\amax > 1 $mm).

However, before fixing upon this conclusion, we will now examine the alternative possibility that there is no significant grain growth to scales $ > 100 \mu$m and that the required spectral index is obtained via mixing the emission from such small dust grains with optically thick substructures. We have shown above that this does not work if the material with high spectral index is placed at large radius because this predicts a steeply decreasing disc size as a function of wavelength. However in this section we consider the case where optically thick and thin material are co-located at all radii. From the point of view of our analysis, it does not matter whether these substructures are large scale coherent structures that would be potentially resolvable with long baseline observations or whether they represent small scale dust condensations that would always remain at the sub-beam level. 

Our purpose therefore is to try and construct a model that circumvents the conclusion that grain growth to\,mm scales in protoplanetary discs is inescapable. In this model we posit a background of {\it small} grains with $\beta = 2$ and a population of optically thick substructures.
The details of the model are given in Appendix~\ref{app:detailed.toymodel.results:structured}.
We prescribe the radial variation of the area filling factor of optically thick substructures and of the optical depth of the medium and generate large suites of models that alter the balance between emission in the two components and the optical depth normalisation for the background. 
Given the degree of flexibility in the models, it is unsurprising that we find some parameters that work; successful models however occupy a narrow niche of parameter space. 
In practice we find that the only models that work are those where (i) there is a roughly equal balance between the emission from optically thick substructures and background emission and (ii) this balance shows little variation across the disc. 
The rough balance between the two emission components is set by the requirement that the integrated spectral index has an intermediate value between the values of 2 and 4 expected for optically thick emission and for small grains in the optically thin limit, respectively. The lack of radial variation in the balance between these components is set by the need to reproduce the similar $\Rse$ values at 0.9 and 3~mm.

\begin{figure}
\centering
\includegraphics[scale=1]{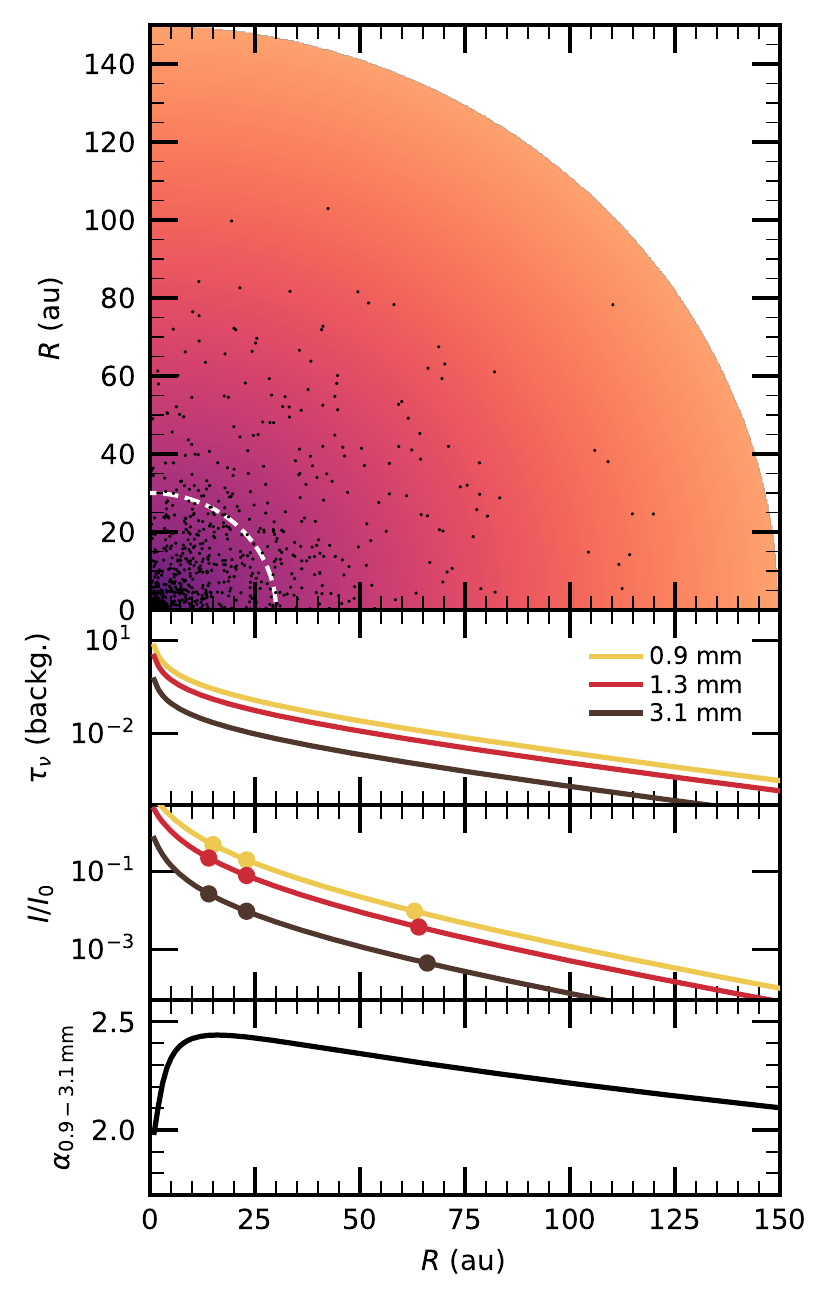}
\caption{
Sketch of a structured toy model occupying the narrow range of parameter space  ($\tau_\mathrm{1.3\,mm}(R_c)=0.04$, $\ff_\mathrm{1.3\,mm}(R_c)=0.04$: see Appendix~\ref{app:detailed.toymodel.results:structured}) that satisfies observational constraints (i.e. $\alphamm=2.35$, $\ff_\mathrm{1.3\,mm}=0.33$).
The top panel shows a quarter of the protoplanetary disc structure. The model assumes a modified self-similar brightness profile ($R_c=30\,$au, $\gamma_1=-1$, $\gamma_2=1$) for the background and optically thick structures depicted as black dots. A dashed white line highlights the scale radius $R_c$. The three panels at the bottom show the optical depth of the background ($\tau_\nu$), the total brightness (including optically thick structures) normalised at the inner radius ($I_\nu/I_0$), and the emerging spectral index profile. Radii enclosing 50, 68, and 95\% of the emission are shown as thick circles in the brightness plot.
}
\label{fig:toymodel.structured.sketch}
\end{figure}

\figref{fig:toymodel.structured.sketch} depicts a model in the small niche of parameter space that
is able to reproduce typical properties of discs in Lupus.
The top panel depicts a quarter of the disc, where the brightness of the background of small grains is represented with the colour scale, and the optically thick substructures as black circles: 
the area covering factor of optically thick structures is $\sim 0.04$ at all radii. We emphasise that it makes no difference to the optical properties whether these optically thick structures are positioned randomly (as shown) or arranged in rings or spirals. The three lower panels represent the background optical depth, the total brightness (including emission from the optically thick structures), and the emerging spectral index profile, respectively.

We conclude from this that if there is no grain growth to $\amax>100 \mu$m, and if the observed spectral indices are explained in terms of optically thick substructures, this can be made to work only under contrived conditions regarding the balance of the emission components. Whereas we of course could not rule out such an interpretation for a particular source, the fact that all the sources would require this particular combination of parameters makes us disfavour this possibility.

We emphasise, however, that our analysis is not disfavouring the possibility of optically thick substructures in general  but that this cannot readily be made to work if the distributed dust component is composed of very \textit{small} grains (with $\beta = 2$). We found that when we modeled a mixture of substructures and  a background composed of large grains ($\beta = 1$), a  wide variety of realisations were broadly compatible with the observed system properties.

\subsection{Summary of constraints derived from the data}
The data is rather tightly clustered around spectral index $\sim 2.4-3$, with a moderate optically thick fraction (most sources in the range $0.2-0.6$) and a ratio of radii at 3.1\,mm to 0.88\,mm that is close to unity. This combination of parameters can be easily realised by smooth disc profiles where the opacity index, $\beta$, is spatially constant and with a value in the range $0.5-1$ in most cases. For constant $\beta$ models, the main constraint is provided by optically thick fraction and spectral index data. Models that match this data automatically satisfy the requirement of having very similar radii in the two wavebands. 

It is also possible to find models where $\beta$ is a smoothly varying function of radius. For example there is a wide range of profiles with low $\beta$ in the interior and large $\beta$ at large radius which can accommodate the observed values of spectral index and optical depth function. Crucially, however, most such solutions predict that the disc is significantly smaller at $3.1$\,mm than at $0.88$\,mm (see the red regions in Fig.~\ref{fig:toy_model.smooth.grid.beta.linear}). This is because if $\beta$ is increasing to high values over regions of the disc producing a significant fraction of the flux, the longer wavelength emission is being down-weighted at large radius, resulting in small $\Rse$ values. We find that the only profiles that are also consistent with the wavelength independence of $\Rse$ are those where the asymptotic value of $\beta$ (namely, $\beta_\mathrm{out}$) is in the range $0.5-1$ and where only the innermost regions ($10-15 \%$ of $\Rse$) can have lower $\beta$ values.  The main constraint on the size of interior regions with low $\beta$ is set by spectral index data, since a large inner region of low $\beta$ would lead to predicted spectral indices that are too low.
We have also explored whether the above constraints on the permissible profiles of $\beta$ necessarily  constrain the microphysical dust properties (see Sect.~\ref{sect:dustprop} below).

Furthermore, we have investigated whether a mixture of optically thick regions and optically thin regions with much higher $\beta$ values ($\sim 2$)  could  be consistent with the data. We conclude that this can match the data only if the ratio of optically thin to  optically thick emission does not vary with radius and is of order unity. In the absence of a reason for expecting such a universal distribution for the mixture of optically thin and thick material, we disfavour this possibility. 

Finally, a further possibility is that the disc emission is dominated by optically thick substructures but that these consist of grains where the albedo is high and increasing towards longer wavelengths. Such substructures can produce emission with spectral index similar to that observed without the need to add an optically thin background and automatically predict that the disc radius is insensitive to wavelength.

It is worth emphasising two further points. First of all our conclusion that the data favours $\beta \sim 0.5-1$  or else optically thick emission from grains with high albedo that is an increasing function of wavelength refers to the bulk of the emission: we cannot rule out the presence of localised (and spatially unresolved) optically thin zones with higher $\beta$ (thus, higher $\alphamm$) provided that they are not a major component of the flux.
Secondly, while we conclude that the data cannot be readily accommodated by a mixture of high $\beta\sim 2$ material and optically thick zones,  the data is readily fit by range of models combining optically thick emission with distributed \textit{low} $\beta \sim 1$ material or by purely optically thick emission with suitable scattering properties.

\subsection{Implications for dust properties}
\label{sect:dustprop}

We have argued above that in the case of smoothly structured discs, $\beta = 0.5-1$
over the bulk of the disc in most sources and that we cannot readily reconcile the data with a mixture of emission from regions with high $\beta$ combined with optically thick substructures. In this case, the value of $\beta$ can be immediately linked to the emission properties of the dust and hence the grain size distribution. 

\begin{figure}
\centering
\includegraphics[scale=1]{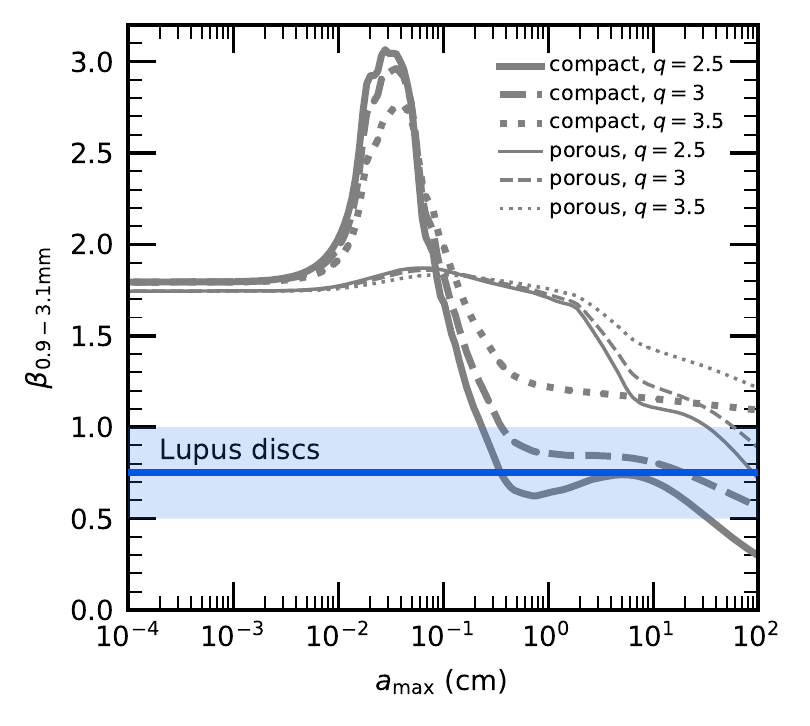}
\caption{Dust opacity spectral index as a function of maximum grain size for grain populations with different size distributions and physical properties. The thick  blue line and light blue uncertainty band represents the typical value and range occupied by the majority of the Lupus data although a few sources are compatible with slightly higher $\beta$ values (up to $\sim 1.5$).}
\label{fig:beta.vs.amax.constraints}
\end{figure}

\figref{fig:beta.vs.amax.constraints} depicts the theoretical value of the dust opacity spectral index $\beta$ between 0.9 and 3.1\,mm as a function of maximum grain size $\amax$ for a variety of assumptions about the grain size distribution and grain porosity. 

The opacity curves were obtained using the \texttt{dsharp\_opac} Python package\footnote{Code available at \href{https://github.com/birnstiel/dsharp_opac}{https://github.com/birnstiel/dsharp\_opac}.} \citep{Birnstiel:2018aa}, which implements a Mie-theory dust opacity model, with appropriate mixing rules for composite materials. We compute the opacity for compact grains (with a composition that is labelled \textit{default} in the DSHARP analysis, see Sect. 2 in \citealt{Birnstiel:2018aa}) and for porous  grains (with same fractional abundances of compact grains, but with 80\% porosity), and for different values of $q=2.5,\ 3,\ 3.5$, with $q$ being the slope of the power-law grain size distribution $n(a)\propto a^{-q}$ for $a_\mathrm{min}\leq a \leq \amax$ ($a$ being the dust particle radius). The curves shown in \figref{fig:beta.vs.amax.constraints} refer to the absorption opacity, as is relevant for interpreting spectral indices of optically thin emission.
The range of values that are consistent with the bulk of our Lupus data are denoted by the blue band. 

\figref{fig:beta.vs.amax.constraints} demonstrate that only compact grains with a relatively top-heavy grain size distribution (grain size power index $q=2.5$ or $3$) are compatible with the bulk of data and then, only if the maximum grain size is in excess of $1$\,mm. Such a maximum grain size is above the grain size corresponding to the opacity resonance feature at all the wavelengths studied. 
The fact that the predicted opacity curves are rather flat for larger maximum 
grain sizes shows that the observed value is compatible with a wide range of maximum grain sizes, consistent with the apparent ubiquity of this $\beta$ value for the Lupus sample.

While top heavy ($q=2.5-3$)  distributions of compact grains with $\amax$ larger than a few mm have a predicted spectral index which is nearly independent of $\amax$, the corresponding opacity value is steeply dependent on $\amax$ in this range. Thus changing $\amax$ from the mm to the cm range reduces the opacity by around an order of magnitude. Thus although our measurements have placed a lower limit on $\amax$, accurate estimation of disc \textit{mass} requires further observations that can distinguish between $\amax$ values in the mm and cm range. In Sect.~\ref{sect:longer.wavelengths.predictions} we discuss how future observations at longer wavelength will be able to distinguish these scenarios.

Models for dust growth in discs \citet[e.g.,][]{Birnstiel:2012yg} predict that grains rapidly grow to scales substantially more than a\,mm but that they are then subject to radial drift which drives the dust to gas ratios down to very low values (e.g. $10^{-4}$ by an age of $3$ Myr). 
Given typical dust masses estimates in protoplanetary discs, such low dust to gas ratios would imply unacceptably large gas masses so it is widely believed that radial drift is somehow inhibited, a popular option being the existence of dust traps in local pressure maxima \citep{Pinilla:2012aa}, possibly associated with the presence of protoplanets. Our demonstration that emission from discs in Lupus is dominated by grains larger than $1$mm (which would otherwise be subject to rapid radial drift) is strong, albeit indirect, evidence that dust trapping must be effective in these discs. 
The resolution of these observations does not allow the detection of sub-structure but some of the objects have been targeted by the DSHARP survey \citep{Andrews:2018aa} which finds evidence of annular structures of various strengths in most of the sources targeted.  

\subsection{Predictions at longer wavelengths}
\label{sect:longer.wavelengths.predictions}
The Q and Ka-Bands of the VLA offer the possibility to observe protoplanetary discs in the wavelength range between 7\,mm and 1\,cm (e.g., with spectral windows centered at 42 and 30\,GHz), which will also be accessible in future via ALMA Band~1 (planned frequency coverage between 35 and 50\,GHz). In future it will be possible to assemble information on disc radii, spectral indices and optical depth fractions for large samples of discs at wavelengths around three times longer than in the present study; to date, however, this information is available for relatively few discs (see \figref{fig:size-frequency.comparison}).
By comparing such new measurements to the disc radii, masses, and spectral indices of discs in different regions or different evolutionary stages (e.g., the Class~0/I objects in the Orion and Perseus clouds; \citealt{Segura-Cox:2018td,Tobin:2020vg,Sheehan:2020ve}) it will be possible to trace the evolution of solids across different stages of disc evolution.

While future datasets should be subject to a modeling exercise similar to that conducted here, it is already possible to make some broad statements about how such measurements could  be used to distinguish between the various possibilities that are compatible with the present dataset.
At ALMA wavelengths, we  have found two types of models that can readily fit the data  i.e. smooth models where the emission is predominantly optically thin  (see Section 6.3.1) and the grain size is in excess of $\sim 1$mm and models where the bulk of the emission derives from optically thick material with a moderate area filling factor see Section 6.4; in the latter case, it is also necessary that the optically thick emission results from grains which are similarly large since such grains have a high scattering albedo which increases towards longer wavelengths and can explain the observed spectral indices.
 
In either scenario, if  the emission is dominated by very large grains (i.e. pebbles on a cm scale or larger), both the absorption opacity and scattering opacity can be described as a single power law over wavelengths ranging  from 1\,mm to 1\,cm \citep{Carrasco-Gonzalez:2019aa}. 
In this case we would  expect little change in the spectral index in this wavelength range and that  the 68th percentile flux radius  would  vary little as a function of wavelength (as in the  grey lines in \figref{fig:size-frequency.comparison}). The smooth variation of optical properties of large grains over this wavelength range would also imply that the optical depth fraction would be slightly lower at 1\,cm than at mm wavelengths,  continuing the trend seen in \figref{fig:ff_vs_Ftot_sample_clean}.

If instead the grain size were towards the lower end of the range allowed by our present modeling (i.e., around a few mm, so in excess of the size corresponding to the opacity resonance at $1-3$\,mm but below the resonant value at cm wavelength), the predictions for the structured models at a wavelength of 1 cm would be somewhat modified, since the spectral index declines towards longer wavelengths for optically thick emission when scattering is included \citep{Zhu:2019aa}. 
However in the case of smoothly structured, largely optically thin models with maximum grain size of a few mm, the spectral index would be expected to rise significantly between mm and cm wavelengths, reflecting the abrupt reduction in opacity in the case that the grains are significantly smaller than the wavelength of emission. This rise in spectral index would produce  a more marked decrease in the optical depth fraction and in the disc size as a function of wavelength.

We therefore conclude that observations at longer wavelengths have the capacity to further constrain the grain size distribution (discriminating between mm and cm scale grains) 
though detailed modeling would be required to firm up these expectations.

\section{Conclusions}
\label{sect:conclusions}
In this paper we have presented the analysis of multi-wavelength ALMA observations at 0.88, 1.3, and 3.1\,mm of 26 protoplanetary discs in the Lupus star forming region. The observations have an angular resolution between 0.25\arcsec and 0.35\arcsec and a comparable sensitivity.

We have derived the multi-wavelength radial brightness profiles of these discs by fitting the interferometric visibilities with parametrised brightness models. 
At each wavelength, we derived the disc effective size ($\Rse$, enclosing 68\% of the disc emission), and the spectral index radial profiles. The homogeneity of the observations (in terms of sensitivity and resolution) and of the analysis enabled us to characterise and compare the properties of all discs with a  minimum relative bias across the wavelengths.

We emphasise that the fact that the discs are spatially resolved at multiple wavelengths (hence we have information on their size) allowed us to break degeneracies in interpreting the spectral index information alone. By forward modelling these observations with simple toy models for the disc emission enables us to present the strongest evidence to date that substantial grain growth (to scales $> 1$\,mm) is required in a large sample of discs, irrespective of their fluxes and radii.

The main results can be summarised as follows:
\begin{enumerate}[(1)]
    \item
    millimeter continuum size-luminosity relation:
    we confirm the relation at 0.9\,mm \citep{Tripathi:2017aa,Andrews:2018ab} and we discover that such relation is present with the same slope when observed at 1.3\,mm, and with a steeper slope when observed at 3.1\,mm, suggesting that large discs are preferentially characterised by a larger $\alphamm$ spectral index.
    \item
    millimeter continuum size-frequency relation:
    in the 0.88-3.1\,mm wavelength range this relation is flat (i.e., $\Rse$ is wavelength-independent), indicating that grains emitting at 0.9-1.3mm ($\amax\sim0.17\,$mm) are essentially co-located. The size-frequency relations found for the Lupus discs are compatible with literature measurement, which typically map the\,mm-brightest objects and lie at the steeper end of the distribution of slopes. This is confirmed by the tentative evidence that the steepness of the millimeter continuum size-frequency relation correlates with the 3\,mm disc luminosity.
    \item
    except for the peculiar case of IM~Lup, our analysis indicates that most of the Lupus discs require large grains ($\amax>1$\,mm) at large radii (50-100\,au), implying that radial drift has to be significantly halted.
    \item
    using the optically thick fraction $\mathcal{F}$ to estimate the amount of emission that can be ascribed to optically thick regions, we prove that Lupus discs are systematically optically thinner at longer wavelengths. 
    Lupus discs are clustered around  optically thick fractions $0.2\leq\ff_\mathrm{1.3mm}\leq0.6$, spectral indices $2.4\leq\alphamm\leq3.0$, and multi-wavelength size ratios $R_\mathrm{68,3.1mm}/R_\mathrm{68,0.9mm}\simeq (0.92\pm0.10)$.
    \item
    by modelling observations with simple models of smooth discs
    we conclude that a ready  way to reproduce the Lupus measurements of $\ff_\mathrm{1.3mm},\alphamm$, and size ratios is to invoke a dust distribution where the opacity index $\beta$ is in the range of $0.5-1$ over a substantial fraction of the disc, although some lower $\beta\sim0$ material at smaller radius is also allowed. 
    \item
    we also model observations in an alternative scenario in which discs are populated by small grains ($\amax<100 \mu$m) and a large number of optically thick substructures: we find that this model can work only under contrived conditions regarding the balance between the emission from the optically thin and thick regions. Although we could not rule out such an interpretation for a particular source, the fact that all the sources would require this particular combination of parameters makes us disfavour this possibility.
    \item a further possible scenario is that the emission is instead entirely dominated by optically thick substructures with a small area filling factor. In this case it is necessary, in order to explain the fact that the spectral indices are significantly larger than $2$, that these substructures are composed of large grains with a high scattering albedo. The constraints on required grain size in this scenario are similar to those for the smoothly structured models described above.
    \item
    in terms of grain growth, the observations of the bulk of the Lupus sample can be explained with $\beta\simeq(0.75\pm0.25)$, which can be produced only by compact grains with a top-heavy grain size distribution ($q=2.5-3$) and $\amax>1$\,mm.
\end{enumerate}

\section*{Acknowledgements}
We thank Richard Booth for several useful discussions and Megan Ansdell for providing the calibrated ALMA datasets at Band~6. We are grateful to the anonymous referee for providing valuable comments that improved the clarity of the paper.
This paper makes use of the following ALMA data: 
ADS/JAO.ALMA\#2016.1.00571.S, 
ADS/JAO.ALMA\#2013.1.00220.S,  
ADS/JAO.ALMA\#2015.1.00222.S, 
ADS/JAO.ALMA\#2013.1.00226.S. 
ALMA is a partnership of ESO (representing its member states), NSF (USA) and NINS (Japan), 
together with NRC (Canada), MOST and ASIAA (Taiwan), and KASI (Republic of Korea), in 
cooperation with the Republic of Chile. The Joint ALMA Observatory is operated by 
ESO, AUI/NRAO and NAOJ.
M.T. and C.J.C. have  been supported by the UK Science and Technology research Council (STFC) via the  consolidated grant ST/S000623/1 and by the European Union’s Horizon 2020 research and innovation programme under the Marie Sklodowska-Curie grant agreement No. 823823 (RISE DUSTBUSTERS project).
J.P.W. acknowledges support from NSF grant AST-1907486.
G.R. acknowledges support from the Netherlands Organisation for Scientific Research (NWO, program number 016.Veni.192.233) and from an STFC Ernest Rutherford Fellowship (grant number ST/T003855/1).
This research was partially supported by the Italian Ministry of Education, Universities and Research through the grant Progetti Premiali 2012 iALMA(CUP C52I13000140001), the Deutsche Forschungsgemeinschaft (DFG, German Research Foundation) - Ref no. FOR 2634/1ER685/11-1 and the DFG cluster of excellence ORIGINS (www.origins-cluster.de), from the EU Horizon 2020 research and innovation programme, Marie Sklodowska-Curie grant agreement 823823 (Dustbusters RISE project), and the European Research Council (ERC) via the ERC Synergy Grant {\em ECOGAL} (grant 855130).
This work was partly supported by the Deutsche Forschungs-Gemeinschaft (DFG, German Research Foundation) - Ref no. FOR 2634/1 TE 1024/1-1.

\section{Data availability}
The raw data used in this article is publicly available on the ALMA Archive (see project codes in the Acknowledgements). \tbref{tb:YSOs}, \tbref{tb:fit.results.ss2s}, and \tbref{tb:fit.results.gaussian} are available in machine-readable format at \href{http://doi.org/10.5281/zenodo.4756381}{http://doi.org/10.5281/zenodo.4756381}.
The data underlying this article will be shared on reasonable request to the corresponding author.


\bibliographystyle{mnras}
\bibliography{mt_disks} 

\pagebreak
\appendix

\section{Millimeter continuum size-integrated flux relation: }
\label{app:size.luminosity.inc.correction}
Here we present the results of the linear regression between the disc size ($\Rse$) and the integrated flux ($F_\nu)$), as opposed to Sect.~\ref{sect:size.luminosity.relations} where we tested the correlation against the disc \textit{face-on} luminosity. The linear regression is now parametrised as:
\begin{equation}
\label{eq:linear.regression.no.inc.corr.def}
\log \left(\frac{\Rse}{\mathrm{au}}\right) = \mathcal{A} +  \mathcal{B} \log\left[ F_\nu\left(\frac{d}{150\,\mathrm{pc}}\right)^2 \right] + \epsilon\,,
\end{equation}
namely, without the $\cos(i)$ re-scaling term in \eqref{eq:linear.regression.def}.
Except from the slightly different re-scaling distance (150 versus 140\,pc), this is the same parametrisation used in \citet{Andrews:2018ab}. 

\begin{figure*}
\centering
\includegraphics[scale=1]{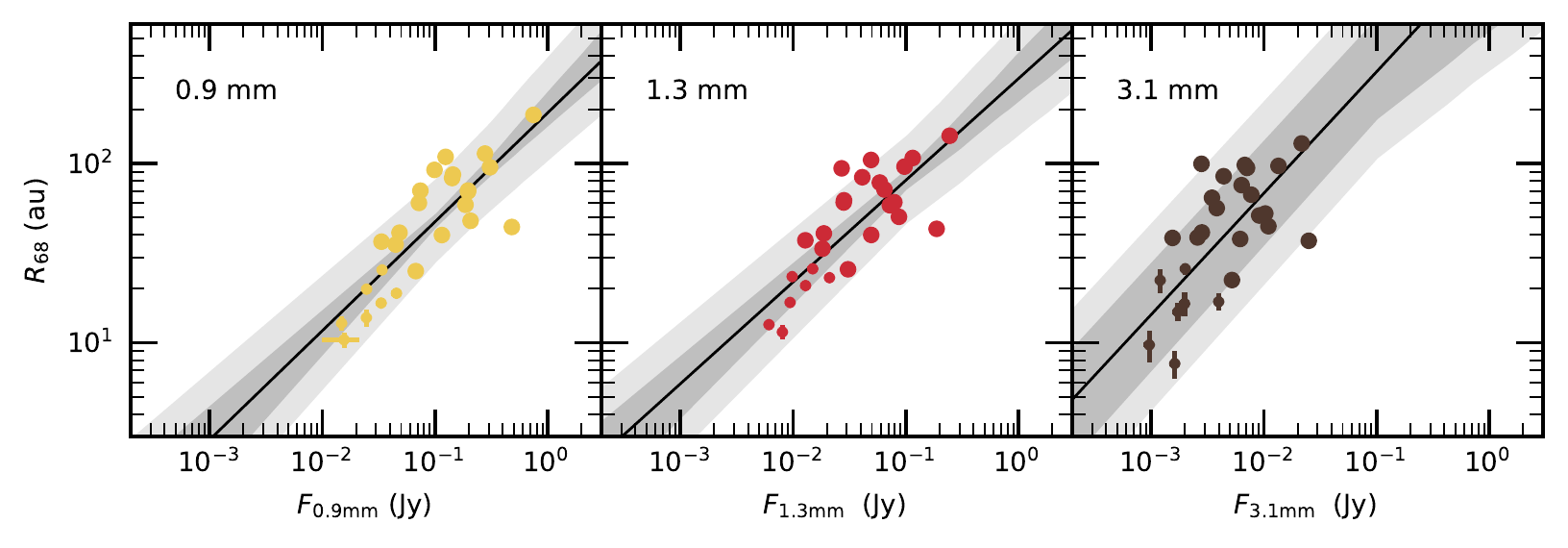}
\caption{Millimeter continuum size - integrated flux relation (Eq.~\ref{eq:linear.regression.no.inc.corr.def}) at multiple wavelengths: 0.9\,mm (left), 1.3\,mm (center), 3.1\,mm (right). The solid black line is the median scaling relation from the Bayesian linear regression. The dark gray area represents the 68\% confidence interval around the median relation, and the light gray area includes the inferred scatter.}
\label{fig:size-luminosity.three.bands.no.inc.corr}
\end{figure*}

By using the same Bayesian linear regression described in Sect.~\ref{sect:size.luminosity.relations} we obtain the results reported in \tbref{tb:linear.regressions.no.inc.corr}. The 0.9 and 1.3\,mm relations are very similar to those found for the \textit{face-on} luminosity (Sect.~\ref{sect:size.luminosity.relations}), while the 3\,mm slope is significantly flatter in this case. Most notably, in these latter linear regressions we find a significantly larger scatter.
\begin{table}
 \caption{Results of the linear regressions for the size-luminosity relation parametrised in Eq.~\eqref{eq:linear.regression.no.inc.corr.def}}
 \centering
 \begingroup
  \setlength{\tabcolsep}{10pt} 
  \renewcommand{\arraystretch}{1.5} 
  \begin{tabular}{lcccc}
   \toprule
$\lambda$ & $\mathcal{A}$ & $\mathcal{B}$ & $\sigma$ & $\rho$ \\ 
\midrule
0.9 mm & ${2.28}^{+0.10}_{-0.10}$ & ${0.61}^{+0.10}_{-0.09}$ & ${0.19}^{+0.04}_{-0.03}$ & ${0.83}^{+0.06}_{-0.09}$ \\ 
1.3 mm & ${2.48}^{+0.15}_{-0.15}$ & ${0.57}^{+0.10}_{-0.10}$ & ${0.19}^{+0.04}_{-0.03}$ & ${0.80}^{+0.07}_{-0.11}$ \\ 
3.1 mm & ${3.21}^{+0.54}_{-0.48}$ & ${0.69}^{+0.24}_{-0.21}$ & ${0.22}^{+0.06}_{-0.05}$ & ${0.76}^{+0.13}_{-0.20}$ \\ 
   \bottomrule
 \end{tabular}
 \endgroup
 \begin{flushleft}
 \textbf{Note.} 
 The values quoted for $\mathcal{A}$ (intercept), $\mathcal{B}$ (slope), $\sigma$ (scatter), and $\rho$ (correlation coefficient)
 are the medians of their posterior distribution; their uncertainties are the central 68\% confidence interval.
 \end{flushleft}
\label{tb:linear.regressions.no.inc.corr}
\end{table}
\figref{fig:size-luminosity.three.bands.no.inc.corr} displays the three correlations with the same colour and line conventions used in \figref{fig:size-luminosity.three.bands}.

\section{Detailed toy model properties: smooth}
\label{app:detailed.toymodel.results:smooth}
\begin{figure*}
\centering
\includegraphics[scale=1, trim={0 1cm 0 0.1cm},clip]{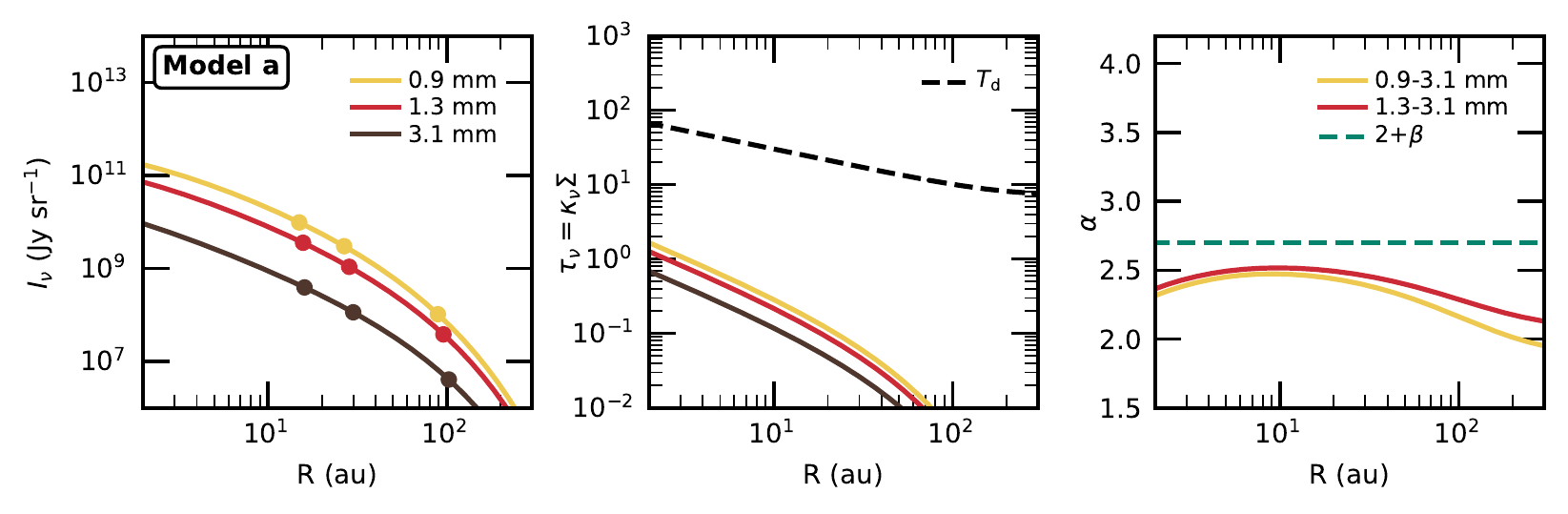}
\includegraphics[scale=1, trim={0 1cm 0 0.1cm},clip]{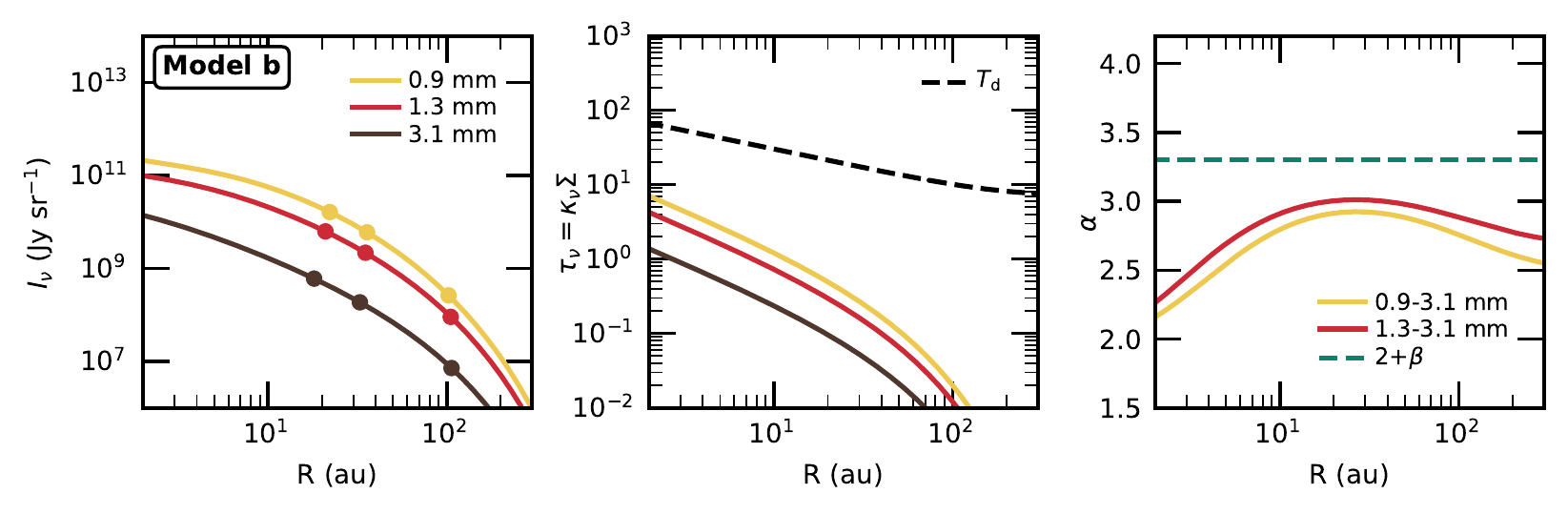}
\includegraphics[scale=1]{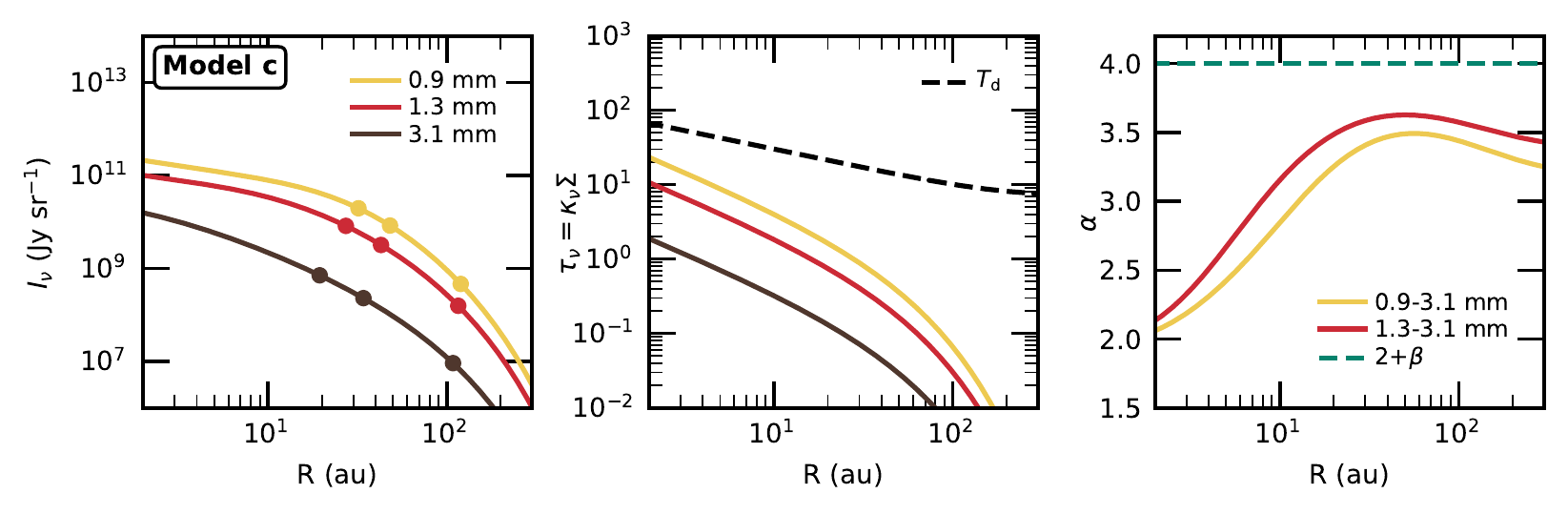}
\caption{
Detailed properties of the  toy model with smooth structure presented in Sect.~\ref{sect:toymodel.smooth}. 
Each of the three panels present detailed properties of the model: brightness profiles (left), optical depth and effective dust temperature (middle), observed spectral index and input dust opacity spectral index (right). The 50\%, 68\%, 95\% flux enclosing radii are highlighted as filled circles in the left panels. 
Model a, b, and c  have been obtained with the following parameters: $\tau_{\nu0}=0.07, 0.3, 1$, and radially uniform $\beta=0.7,\ 1.3,\ 2$, respectively. They all assume $R_c=30$\,au, $\gamma_1=-1$, $\gamma_2=1$.
}
\label{fig:toy_model.smooth.beta.const.appendix}
\end{figure*}

Here we present the detailed properties of the toy model with smooth structure used in Sect.~\ref{sect:toymodel.smooth}. 
To document the behaviour of the model, in \figref{fig:toy_model.smooth.beta.const.appendix} we show three models (a, b, and c), that are representative of different regimes. The three models have different input $\tau_{\nu,0}$ and $\beta$, which allows us to reproduce different disc-integrated values of $\ff_\mathrm{1.3\,mm}$ and $\alphamm$. All the models have radially uniform $\beta$ and assume $R_c=30$\,au, $\gamma_1=-1$, $\gamma_2=1$. 
For each model we present detailed properties: brightness profiles, optical depth and effective dust temperature, observed spectral index and input dust opacity spectral index. The 50\%, 68\%, 95\% flux enclosing radii are highlighted as filled circles in the brightness profiles. 

Model (a) has $\tau_{\nu,0}=0.07$ and $\beta=0.7$: the low opacity spectral index produces slowly varying $\alphamm(R)$ profile, resulting in a disc radius that is essentially constant across 0.9-3.1\,mm wavelength range. Model (b) has higher $\beta=1.3$ and larger variations in $\alphamm(R)$: its 3.1\,mm radius is $\sim$90\% the 0.9\,mm radius (as typically observed for the Lupus discs).
Model (c) has a very large $\beta=2$, which makes $\alphamm(R)$ reach large values $\sim3.5$: this produces a strong reduction of the disc size from 0.9 to 3.1\,mm.

\begin{figure}
\centering
\includegraphics[scale=1]{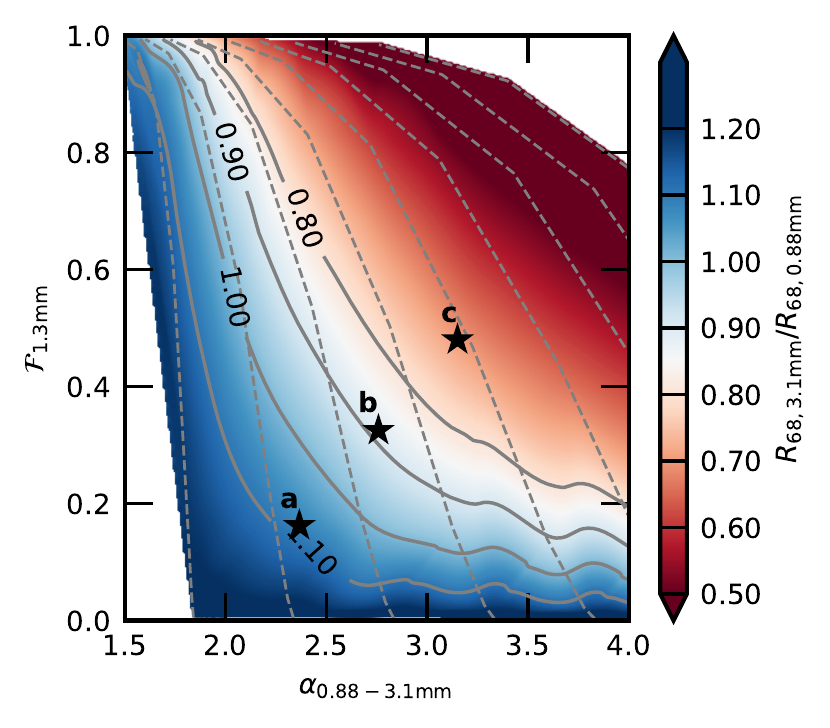}
\caption{
The three toy models (a, b, and c) located at the following coordinates ($\alphamm$, $\ff_\mathrm{1.3\,mm}$): 
$(2.37, 0.16),\ 
(2.76, 0.33),\ 
(3.15, 0.48)$, 
respectively and highlighted in the $\ff_\mathrm{1.3\,mm}-\alphamm$ plane.}
\label{fig:toy_model.smooth.beta.const.appendix.ffalphaplane}
\end{figure}

The location of these three models in the same $\ff_\mathrm{1.3\,mm}-\alphamm$ plane as in \figref{fig:toy_model.smooth.grid.beta.const} are highlighted in \figref{fig:toy_model.smooth.beta.const.appendix.ffalphaplane}.

\section{Detailed toy model properties: structured}
\label{app:detailed.toymodel.results:structured}

In structured models we posit a background of {\it small}  grains  ($< 100 \mu$m) with optical depth profile given by (\ref{eqn:tauprescript}) with  $\beta = 2$. We then introduce the quantity $\xi(R) = X \tau(R)$ (where $X$ is a scaling factor that can in principle be a function of radius) such that the fraction of the disc area at a given location that is occupied by optically thick substructure is given by $1-\exp[-\xi(R)]$. The total emission at that location is then the sum of that from the optically thick substructures and the background of small grains (which occupy an area filling factor of $\exp[-\xi(R)]$). In regions of the disc where the background of the disc is optically thin, the composite emission has  radially constant $\alphamm$ if $X$ is independent of radius; the  value of $\alphamm$ is controlled by the scaling factor $X$ which is the model parameter that controls the relative dominance of emission from the substructures and from the background of small grains.

With this prescription we can generate large suites of models that alter the balance between emission in the two components and  the optical depth normalisation for the background;  in order to find a model that predicts similar radii at 0.9 and 3.1\,mm we consider the case of radially constant $X$ since these models have little radial variation of $\alphamm$. Given the degree of flexibility in the models, it is unsurprising that we find some parameters that work; successful models however occupy a narrow niche of parameter space. Not only is it necessary to invoke radially constant $X$ (as noted above) but the value of  $X$ has to be of order unity, since otherwise the emission would be dominated by one of the components and the resulting spectral index would be either  too high (close to 4 if $X$ is too low) or too low (close to 2 if $X$ is too high).

\section{The role of scattering}
\label{app:role.of.scattering}
The toy-model analysis \textbf{we presented in Section~\ref{sect:discussion}} relied on considering thermal emission without scattering and a prescribed frequency dependence for the optical depth associated with absorption (Eq.~\ref{eqn:tauprescript}). Inclusion of scattering opacity has no effect on the emission properties at low optical depth. However, in optically thick regions,  scattering {\it{reduces}} the emission below its black body value: the increased path length of photons undergoing multiple scattering means that the effective surface of last emission moves higher in the disc atmosphere, implying that radiation  derives from a smaller column of emitting material \citep{Carrasco-Gonzalez:2019aa,Zhu:2019aa}. 
The effect on the spectral index from optically thick regions depends on the wavelength dependence of the albedo which can have either sign. Consequently, the canonical value of $\alpha=2$ for optically thick emission in the Rayleigh Jeans limit can be modified by scattering to  lie in the range $\sim 1.6-2.5$ \citep{Zhu:2019aa}.

These considerations do not significantly modify the conclusions from the modelling Sections ~\ref{sect:toymodel.smooth},\ref{sect:toymodel.structured}. In the case of the smooth radial profiles considered in Section~\ref{sect:toymodel.smooth}, the models that match the rather low optically thick fraction values of much of the  data contain only a minor contribution from optically thick emission and would therefore not be affected by inclusion of scattering. 
The models considered in Section~\ref{sect:toymodel.structured} contained only small grains ($< 100\,\mu$m) for which the albedo is very low and thus would not be affected by scattering even within the optically thick substructures.

Consideration of scattering however permits another interpretation of the distribution of the observational data in the plane of spectral index versus optically thick fraction. If the albedo {\it rises} with wavelength in the range $0.88-3.1$mm, then this suppresses the flux at 3\,mm in regions of high optical depth, i.e. it increases the spectral index in optically thick regions to $>2$. 
For sufficiently large grains, $\alphamm$ can attain a value of $2.3-2.5$ \citep{Zhu:2019aa}, which is typical of the observed values in our sample. In this case, the data would be consistent with  emission deriving almost exclusively from optically thick structures composed of large grains, with the observed optically thick fraction requiring an area filling factor of optically thick emission of tens of percent. 
Such a scenario would automatically satisfy the requirement of having the same radii at different mm wavebands, since emission in each of these bands would derive from the same structures. \citet{Carrasco-Gonzalez:2019aa} successfully model the resolved multi-wavelength profiles of HL~Tau with such a scenario (where in this case the optical depth at 1.3 mm is at least modestly larger than unity area over a large radial range).

\section{Detailed fit results}
\label{app:detailed.fit.results}
In this Appendix we report the detailed fit results for the 26 discs that we considered for the multi-wavelength visibility modelling (Sect.~\ref{sect:analysis}). In Sect.~\ref{app:detailed.fit.results.selfsimilar} we present the fits performed with the self-similar brightness profile, and in Sect.~\ref{app:detailed.fit.results.gaussian} the fits performed with the Gaussian profile.
\ \\
\ \\
\textit{(This appendix is available online as supplementary material)}

\subsection{Fits with modified self-similar profile}
\label{app:detailed.fit.results.selfsimilar}

\subsection{Fits with Gaussian profile}
\label{app:detailed.fit.results.gaussian}

\bsp  
\label{lastpage}
\end{document}